%% file: main-arXiv.tex
\documentclass[12pt,a4paper,table]{article}

\usepackage{floatrow}
\usepackage{relsize}
\input{preamble}

\input{belle2-symbols.tex}

\graphicspath{{figures/}}

\usepackage{fancyhdr}
\fancypagestyle{title}{%
    \setlength{\headheight}{3\baselineskip}%
    \fancyhead[R]{Belle II Preprint 2026-009 \\ KEK Preprint 2026-5\\[\baselineskip]}%
}

\title{Charged-lepton identification at \BelleII}
\date{}

\begin{document}

\pagenumbering{gobble}


\maketitle

\noindent
\input{authors-orcid.tex}

\begin{abstract}
    \input{0_abstract}
\end{abstract}

\newpage


\pagenumbering{arabic}

\input{body}


\input{acknowledgements-b2}

\addcontentsline{toc}{section}{References}
\bibliographystyle{belle2-note}
\bibliography{belle2}

\end{document}

%% file: preamble.tex
\usepackage{hyperref}    
\usepackage{orcidlink}
\usepackage[italic]{belle2-particles}
\usepackage{belle2-symbols}
\usepackage{amsfonts}
\usepackage{amsmath} 
\usepackage{amssymb}
\usepackage{array}
\usepackage{blindtext} 
\usepackage{bm}
\usepackage{caption}
\usepackage{cite} 
\usepackage[noabbrev,capitalise]{cleveref}
\usepackage{color}
\usepackage{colortbl}
\usepackage{floatrow}
\usepackage{hypcap} 
\usepackage{ifpdf}
\usepackage{graphicx}  
\usepackage{highlight} 
\usepackage{longtable}
\usepackage{lineno}  
\usepackage{mathrsfs}
\usepackage{mciteplus}
\usepackage[version=4]{mhchem}
\usepackage{microtype}
\usepackage{multirow}
\usepackage{overpic}
\usepackage{pgffor} 
\usepackage{physics}
\usepackage{rotating}
\usepackage[separate-uncertainty]{siunitx} 
\usepackage{subfigure}
\usepackage{tikz}
\usepackage{xfp} 
\usepackage{xspace} 
\usepackage{upgreek} 
\usepackage{url}

\usepackage{forloop}

\graphicspath{{./figs/}} 

\newcommand*\patchAmsMathEnvironmentForLineno[1]{%
\expandafter\let\csname old#1\expandafter\endcsname\csname #1\endcsname
\expandafter\let\csname oldend#1\expandafter\endcsname\csname
end#1\endcsname
 \renewenvironment{#1}%
   {\linenomath\csname old#1\endcsname}%
   {\csname oldend#1\endcsname\endlinenomath}%
}
\newcommand*\patchBothAmsMathEnvironmentsForLineno[1]{%
  \patchAmsMathEnvironmentForLineno{#1}%
  \patchAmsMathEnvironmentForLineno{#1*}%
}
\AtBeginDocument{%
\patchBothAmsMathEnvironmentsForLineno{equation}%
\patchBothAmsMathEnvironmentsForLineno{align}%
\patchBothAmsMathEnvironmentsForLineno{flalign}%
\patchBothAmsMathEnvironmentsForLineno{alignat}%
\patchBothAmsMathEnvironmentsForLineno{gather}%
\patchBothAmsMathEnvironmentsForLineno{multline}%
}

\newfloatcommand{capbtabbox}{table}[][\FBwidth]

\ExplSyntaxOn
\RenewDocumentCommand \SIrange { O{} m m m } {
  \leavevmode
  \group_begin:
     \keys_set:nn { siunitx } { open-bracket = [,  close-bracket= ], #1}
     \__siunitx_range_unit:nnnn {#4} { open-bracket = [,  close-bracket= ], #1} {#2} {#3}
  \group_end:
}
\ExplSyntaxOff

\NewDocumentCommand{\dEdx}{s}{\ensuremath{\IfBooleanTF{#1}{\dv*{E}{x}}{\dv{E}{x}}}\xspace}

\usepackage{setspace}

%% file: authors-orcid.tex
  \author{M.~Abumusabh\,\orcidlink{0009-0004-1031-5425}} 
  \author{I.~Adachi\,\orcidlink{0000-0003-2287-0173}} 
  \author{A.~Aggarwal\,\orcidlink{0000-0002-5623-3896}} 
  \author{H.~Ahmed\,\orcidlink{0000-0003-3976-7498}} 
  \author{Y.~Ahn\,\orcidlink{0000-0001-6820-0576}} 
  \author{H.~Aihara\,\orcidlink{0000-0002-1907-5964}} 
  \author{M.~Akdag\,\orcidlink{0009-0004-3728-1077}} 
  \author{N.~Akopov\,\orcidlink{0000-0002-4425-2096}} 
  \author{S.~Alghamdi\,\orcidlink{0000-0001-7609-112X}} 
  \author{M.~Alhakami\,\orcidlink{0000-0002-2234-8628}} 
% ...
{\centering (The Belle II Collaboration)}

%% file: 0_abstract.tex
Effective particle identification capabilities are a strategic priority for the physics program of the \BelleII experiment.
We describe the algorithms used at \BelleII for identifying electrons and muons and separating them from charged hadrons.
We present the performance obtained by the experiment during Run 1, which consists of \SI{428}{fb^{-1}} of data collected at the energy-asymmetric $\APelectron\Pelectron$ collider SuperKEKB between 2019 and 2022 at center-of-mass energies near the mass of the \PUpsilonFourS.

%% file: body.tex
\section*{Introduction}
\label{sec:introduction}

The \BelleII experiment operates at the asymmetric-energy \epem\ collider SuperKEKB~\cite{Akai:2018mbz}, located in Tsukuba, Japan.
Its broad physics program includes the study of \CP violation in $B$- and $D$-meson decays, the precise measurement of the elements of the Cabibbo-Kobayashi-Maskawa quark-mixing matrix, the study of $\tau$-lepton decays, and searches for phenomena that are forbidden or extremely suppressed in the standard model (SM) of particle physics, such as lepton flavor universality violation. In order to achieve its goals, excellent particle identification (PID) capabilities are required. At \BelleII, six long-lived charged particle species typically traverse the entire detector: electrons, muons, pions, kaons, protons, and deuterons.
The ability to distinguish among these species is important not only for effectively identifying the target final states of specific analyses, but also for ensuring optimal performance of analysis tools of general use, such as the $B$-flavor tagger~\cite{Belle-II:2021zvj, Belle-II:2024lwr}, the charm flavor tagger~\cite{Belle-II:2023vra}, and the full reconstruction of $B$ mesons via hadronic or semileptonic decays~\cite{Keck:2018lcd}.

\BelleII expands the physics program of the \babar and Belle experiments~\cite{Bevan:2014iga}, which in the first decade of this century accumulated a combined integrated luminosity of $\sim 1.5$ ab$^{-1}$ at a center-of-mass (c.m.) energy corresponding to, or near, the mass of the \FourS resonance. \BelleII aims at integrating 50 ab$^{-1}$, at similar energies. The required, much higher instantaneous luminosity poses significant challenges to both the hardware and the reconstruction software of the experiment, due to the harsher background conditions in which they have to operate. The first period of physics data taking (the so-called Run 1) began in the year 2019 and was concluded in 2022, after integrating 428 fb$^{-1}$. The experiment then stopped for one and a half years to perform maintenance and upgrade work, before resuming data taking in 2024. 

The results presented in this paper are based on the Run 1 data set. We recently summarized the performance of hadron ($\pi$, $K$, $p$) identification at \BelleII in Ref.~\cite{Belle-II:2025tpe}. In this contribution we will focus on the identification of charged leptons ($e$ and $\mu$) and on our capability of distinguishing them from hadrons (since electrons and muons have very different behaviors in our detector, $e$-as-$\mu$ and $\mu$-as-$e$ mis-identification rates are negligible in most cases). 
The paper is organized as follows. In sections \ref{sec:detector} and \ref{sec:data}, we provide a brief overview of the \BelleII detector and of the samples that will be utilized for the performance studies. We then describe how each subdetector provides PID information in section \ref{sec:likelihoods} and how this information is combined in variables that are easy to apply by nonexperts in physics analysis in section \ref{sec:probabilities}. In sections \ref{sec:eff_and_misID} and \ref{sec:performance} we define the performance indicators for PID and present some examples of the typical performance that can be achieved in a \BelleII analysis. We conclude in section \ref{sec:conclusions} with some prospects for the future of the experiment.

\section{Detector}
\label{sec:detector}

\BelleII is a general-purpose detector consisting of seven subdetectors and a superconducting solenoid arranged cylindrically around the \epem interaction region~\cite{Belle-II:2018jsg, Abe:2010sj}. From innermost to outermost, these subdetectors are the pixel vertex detector~(PXD), silicon vertex detector~(SVD), central drift chamber~(CDC), time-of-propagation detector~(TOP) and aerogel ring-imaging Cherenkov detector~(ARICH), electromagnetic calorimeter~(ECL), and \KL and muon detector~(KLM). The solenoid, located between the ECL and the KLM, provides a \SI{1.5}{T} magnetic field nearly parallel to the beam directions. 
We define $\hat{z}$ as the cylindrical axis of the solenoid, with its positive direction nearly coincident with electron-beam direction (the beams collide with a crossing angle of \SI{83}{mrad}). Polar angles are defined relative to $\hat{z}$, and azimuths relative to the direction orthogonal to $\hat{z}$ that points outside the accelerator ring. The origin of the coordinate system is the nominal point at which the beams collide. The actual \epem interaction point depends on the data-taking period and is determined from $e^+e^- \to \mu^+\mu^-$ events. It is typically within a millimeter of the origin.

The PXD~\cite{MOSER201685} consists of two layers of DEPFET (DEpleted P-channel Field Effect Transistor) pixel sensors, the first covering the full azimuthal range and the second only 20\% (a new PXD with a full second layer was installed in 2023).
The SVD~\cite{Adamczyk_2022} consists of four layers of double-sided silicon strip sensors.
The CDC, the main tracking device, is a large volume of helium and ethane gas crossed by sense and field wires.
The TOP~\cite{Atmacan:2025jmh}, covering the barrel region, and the ARICH, covering the forward end cap, measure Cherenkov light produced by charged particles.
The TOP consists of quartz bars which internally reflect light, that is then detected by micro-channel-plate photo-multiplier tubes.
The ARICH~\cite{Yonenaga:2020eby} consists of two layers of aerogel tiles, with different refractive indices, that focus light into sharp rings, detected by hybrid avalanche photon detectors.
The ECL~\cite{Kuzmin:2020new} covers the barrel and forward and backward end caps with thallium-doped cesium-iodide crystals, each \num{16.2} radiation lengths deep.
The KLM~\cite{Ketter:2025yqm} serves as the return yoke of the magnetic field, with gaps in the steel structure instrumented with scintillator strips in the end caps and first two layers of the barrel and with resistive plate chambers in the other barrel layers.
The trigger system of \BelleII consists of an FPGA-based level one trigger (utilizing information from the CDC, ECL, and KLM) and a software-based high-level trigger, exploiting the properties of events reconstructed from all subdetectors (except the PXD).

Our PID uses likelihoods that depend on a particle's momentum, $\vec{p}$, measured from the curvature of the particle's track in the magnetic field, which is reconstructed from the locations of hits in the PXD, SVD, and CDC;
$p$, $\theta$, and $\phi$ are the magnitude, polar angle, and azimuth of this momentum, respectively.
When a reference frame is not explicitly mentioned, variables are defined in the laboratory frame.

\section{Data and simulation}
\label{sec:data}

We use data collected in Run 1 and simulation samples that resemble them to develop and study the PID likelihoods and their performance. We simulate the detectors taking into account the variation of their conditions throughout data taking and overlay background signals taken from pseudo-randomly triggered (i.e. recorded at a fixed time following a collision selected by a physics trigger) events. 
We simulate the production of quark-antiquark, $\mu^+\mu^-$, and $\tau^+\tau^-$ pairs from the \epem collision with \textsc{KKMC}~\cite{Jadach:1999vf}, hadronization with \textsc{PYTHIA~8}~\cite{Sjostrand:2014zea}, hadron decay with \textsc{EvtGen}~\cite{Lange:2001uf}, and \Ptau decay with \textsc{Tauola}~\cite{Davidson:2010rw}. We use \textsc{BabaYaga@NLO}~\cite{Balossini:2008xr} to generate (radiative) Bhabha processes and \textsc{AAFH}~\cite{Berends:1984gf} for two-photon-fusion processes with four leptons in the final state. 
For each charged-particle species, we also generate simulated data with particles isotropically distributed in the detector and evenly distributed in the range of momenta produced at \BelleII. Detector response is simulated with \textsc{GEANT4}~\cite{GEANT4:2002zbu}, and we reconstruct both real and simulated data using the Belle II analysis software framework, \basfii~\cite{Kuhr:2018lps,basf2-zenodo}.

We develop, test, and measure the performance of our PID algorithms using control samples of tracks for which no PID selection criteria have been applied. This is achieved by using samples in which the particle species is known from the context of the decay of a known parent particle or by the tag-and-probe technique, as will be further detailed in the following subsections.

\subsection{$\epem \to \epem (\gamma)$}
\label{subsec:bhabha}

A clean sample of high momentum electrons and positrons can be obtained from the abundant $\epem \to \epem \; (\gamma)$ process, also referred to as (radiative) Bhabha scattering. The tag-and-probe technique is applied for this sample: in order to increase the purity, PID selection criteria are imposed to only one of the tracks (the \emph{tag}), while the other (the \emph{probe}) is utilized to measure in an unbiased way the performance of our algorithms. 

Candidate Bhabha events are recorded using a low-multiplicity trigger, which requires an ECL cluster with an energy above $2\,\gev$ in the barrel region. In order to minimize any bias that the trigger could introduce to the identification efficiency, the events are required to be triggered by the tag particle. Furthermore, all candidates are required to contain exactly two tracks originating from near the interaction point: the distance between the point of closest approach and the interaction point in the $z$ direction ($dz$) must be less than 5 cm, and across the transverse plane ($dr$) must be less than 2 cm. The global likelihood ratio for the electron hypothesis (see Sec.~\ref{sec:probabilities}) for the tag track is required to be greater than 0.95. To suppress background events with missing particles, the squared mass of the system recoiling against the reconstructed \epem pair 
\begin{equation}
M^2_{\rm recoil} = \left( p_{e^+}^{\rm beam} + p_{e^-}^{\rm beam} -  p_{\rm tag} - p_{\rm probe} \right)^2 
\end{equation}
(where $p_{e^{\pm}}^{\rm beam}$ are the four-momenta of the colliding particles and $p_{\rm tag/probe}$ are the four-momenta of the reconstructed particles) is required to be less than $10\,\gevgevcccc$. We do not reconstruct the photon emitted in a radiative decay.

\begin{figure*}[htbp]
    \begin{center}
        \includegraphics[width=0.7\textwidth]{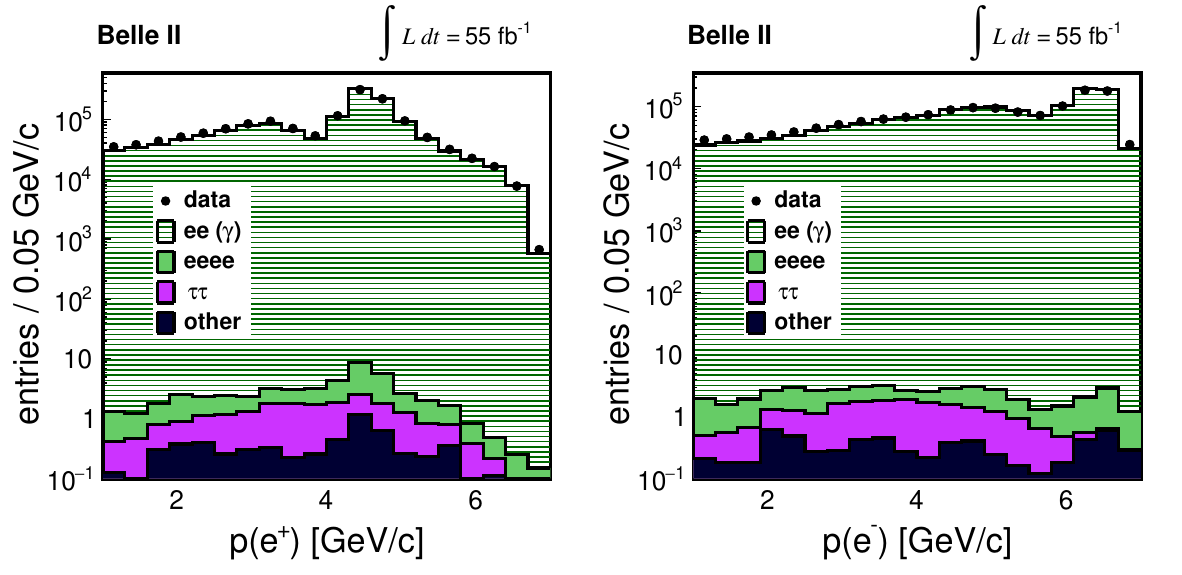}
        \caption{Laboratory-frame momentum distributions of probe positrons (left) and electrons (right) from the radiative Bhabha sample in data (points) and simulation (filled histograms) after all selection criteria, including those on the PID of the tag track, have been applied. The resulting distributions depend on the cross section of the process, on the acceptance of our detector, and on the selection criteria (including trigger requirements) that are applied to the sample.}
        \label{fig:Bhabha_momentum}
    \end{center}
\end{figure*}

Figure~\ref{fig:Bhabha_momentum} shows the momentum distribution of candidate $e^+$ and $e^-$ tracks after the selection (also on the PID of the tag track), obtained from a fraction of the Run 1 data set. The contamination from non-Bhabha processes, estimated from the cross-sections, angular distributions, and simulated reconstruction efficiencies, is of the order of $10^{-4}$ across the whole momentum range. Backgrounds (often also containing actual electrons and positrons) arise from $\epem \to \tau^+\tau^-$, $\epem\epem$, $\epem \mu^+\mu^-$, and $\mu^+\mu^-$ events.

\subsection{$\epem \to \mu^+\mu^- \gamma$}
\label{subsec:dimuon}

\begin{figure*}[htbp]
    \begin{center}
        \includegraphics[width=0.7\textwidth]{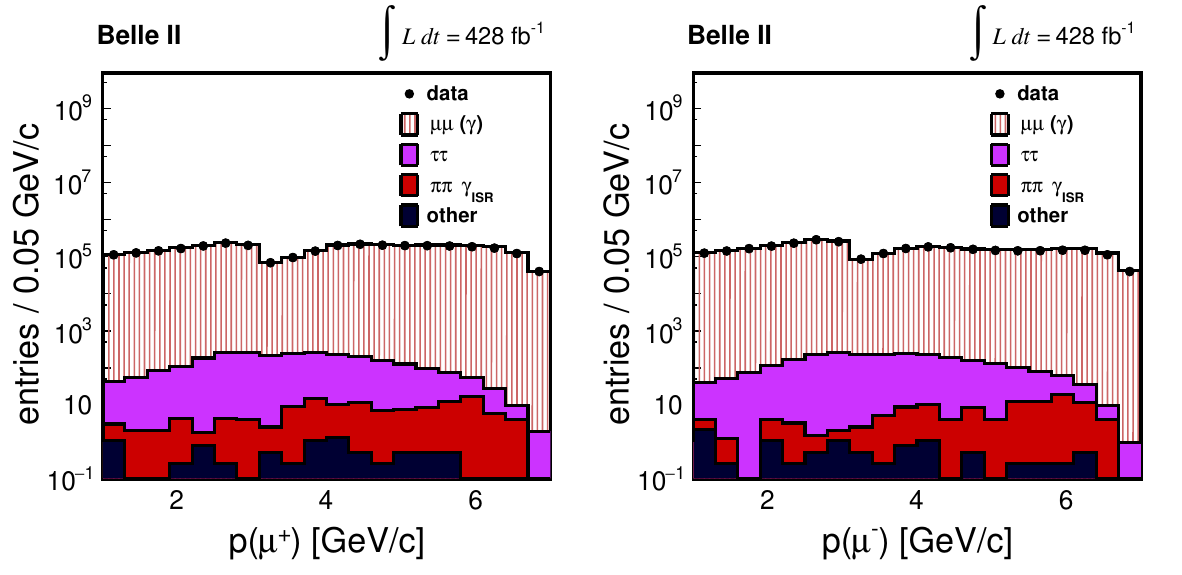}
        \caption{Laboratory-frame momentum distributions of probe $\mu^+$ (left) and $\mu^-$ (right) for the radiative dimuon sample in data (points) and simulation (filled histograms) after all selection criteria, including those on the PID of the tag track, have been applied. The step around $3\,\gevc$ is caused by the high-level trigger selection.}
        \label{fig:Dimuon_momentum}
    \end{center}
\end{figure*}

We use $\epem \to \mu^+\mu^- \gamma$ events to obtain, with the tag-and-probe technique, a high-purity sample of high-momentum muons. Candidates are selected by requiring exactly two oppositely charged tracks, which must originate near the interaction point ($|dz| < 5$ cm, $|dr| < 2$ cm). The photon candidate must be in the ECL acceptance, have a minimum energy of $1.0\,\gev$, and the associated ECL cluster must receive contributions from more than one crystal. We further require the invariant mass of the $\mu^+\mu^- \gamma$ system to satisfy $10.2 < M_{\mu^+\mu^-\gamma} < 10.8\,\gevcc$. Backgrounds arise from the following sources: $\epem \to \tau^+\tau^-$, production of pairs of charged hadrons in events with initial-state radiation ($\epem \to h^+ h^- \, (\pi^0) \, \gamma_{\rm ISR}$),  and two-photon-fusion production of pairs of leptons or hadrons (such as $\epem \to \epem \mu^+ \mu^-$, $\epem \tau^+ \tau^-$, and $\epem \pi^+ \pi^-$). To measure the PID performance, we utilize the tag-and-probe technique, requiring the global likelihood ratio (see Sec.~\ref{sec:probabilities}) for the muon hypothesis of the tag track to be greater than 0.9.
The level of background contamination of this sample is at the percent level or less, as can be seen from Fig.~\ref{fig:Dimuon_momentum}, which shows the momentum of the probe $\mu^+$ and $\mu^-$ for data and simulation. 

The $\epem \to \ell^+ \ell^- (\gamma)$ samples are very large and pure, and cover a very wide momentum range, but exhibit a strong correlation between the magnitude of the momentum and the polar angle of the candidate tracks, such that only a relatively small region of the phase space is densely populated. This limitation is not present in the control samples that we describe in the following.

\subsection{$\epem \to \epem \ell^+\ell^-$}
\label{subsec:diphoton}

We obtain a sample of low-momentum electrons and muons by selecting the two-photon-fusion topology $\epem \to \epem \ell^+ \ell^-$ ($\ell = e, \mu$). The electron and positron from the colliding beams exchange virtual photons that produce a charged lepton pair that is detected; the beam electron and positron, which are only slightly deflected by the interaction, continue down the beam pipe undetected. We trigger the recording of such events with a purely track-based trigger, requiring the presence of two tracks that originate from the interaction region and are back-to-back in the transverse plane. We do not trigger on ECL information because that could bias PID performance. We select events with only two charged tracks, oppositely charged, each with momentum greater than $0.4\,\gevc$ and $|dz| < 5$ cm, $|dr| < 2$ cm. To suppress cosmic muons that are wrongly reconstructed as two tracks, we require the opening angle (in three dimensions) of the tracks to be less than $176^{\circ}$. In the c.m. frame, we require that the sum of the energies of charged and neutral particles detected in the event be less than $6.0\,\gev$ and that the track pair have longitudinal momentum less than $1.0\,\gevc$, transverse momentum less than $0.15\,\gevc$, and mass less than $3.0\,\gevcc$. We employ the tag-and-probe technique, requiring the global likelihood ratio (see Sec.~\ref{sec:probabilities}) for the electron (muon) hypothesis of the tag track to be greater than 0.95.

\begin{figure*}[htbp]
    \begin{center}
        \includegraphics[width=0.7\textwidth]{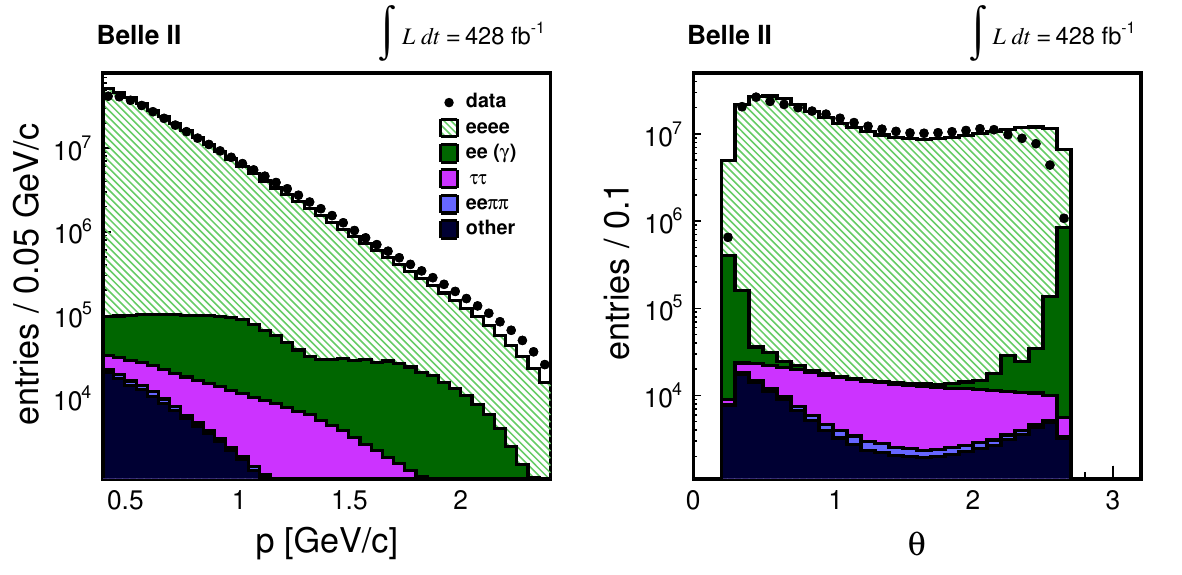}
        \caption{Distributions of the momentum (left) and polar angle (right) for the probe $e^{\pm}$ tracks of the $\epem \to \epem \epem$ sample in data (points) and simulation (filled histograms). All selection criteria, including those on the PID of the tag track, have been applied.} 
        \label{fig:Diphoton_p_theta_e}
    \end{center}
\end{figure*}

\begin{figure*}[htbp]
    \begin{center}
        \includegraphics[width=0.7\textwidth]{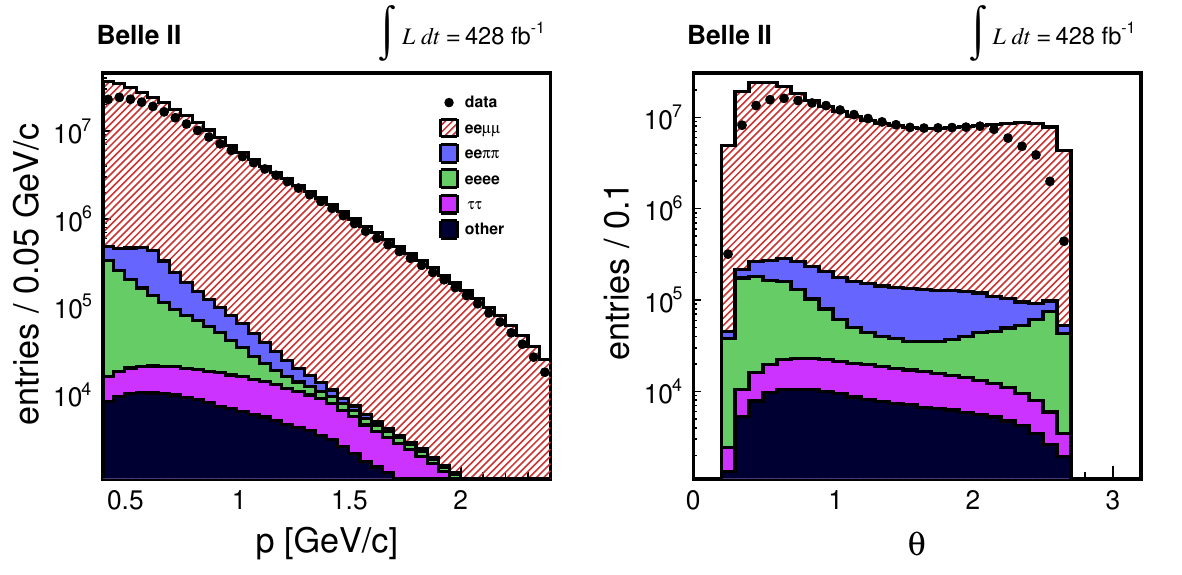}
        \caption{Distributions of the momentum (left) and polar angle (right) for the probe $\mu^{\pm}$ tracks of the $\epem \to \epem \mu^+ \mu^-$ sample in data (points) and simulation (filled histograms). All selection criteria, including those on the PID of the tag track, have been applied.} 
        \label{fig:Diphoton_p_theta_mu}
    \end{center}
\end{figure*}

Figures~\ref{fig:Diphoton_p_theta_e} and \ref{fig:Diphoton_p_theta_mu} show the momentum and $\theta$ distributions for the $\epem \to \epem \epem$ and $\epem \to \epem \mu^+\mu^-$ samples in data and simulation. Backgrounds arise from $\epem \to \epem h^+h^-$, $\epem \to h^+h^- \gamma_{\rm ISR}$ ($h = \pi, K$), and $\epem \to \tau^+\tau^-$ processes. Significant discrepancies between data and simulation arise from the imperfect modeling of the trigger efficiencies in the simulation, and from differences between data and simulation in the PID selection of the tag track. In order to use these samples in the measurement of the PID efficiency, we rescale them (before applying the PID selection on the tag) by the observed data/simulation ratio in each subregion of momentum and $\theta$ and correct for the measured differences between data and simulation of the probabilities of mis-identifying a pion as a lepton (see Sec.~\ref{sec:eff_and_misID}).

\subsection{$J/\psi \to \ell^+\ell^-$}
\label{subsec:Jpsiamples}

\begin{figure*}[htbp]
    \begin{center}
        \includegraphics[width=0.7\textwidth]{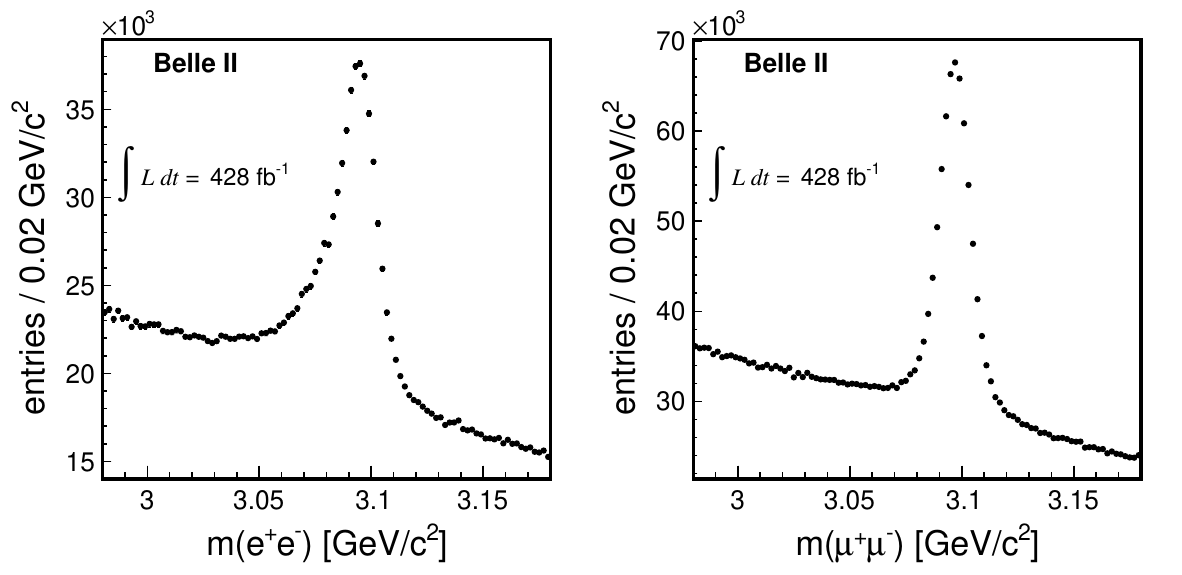}
        \caption{Invariant mass distribution for the $J/\psi \to e^+e^-$ (left) and $J/\psi \to \mu^+\mu^-$ (right) candidates in which at least one of the tracks passes the tag PID selection.} 
        \label{fig:Jpsi_masses}
    \end{center}
\end{figure*}

An inclusive $J/\psi$ sample, with $J/\psi \to \ell^+\ell^-$, provides samples of electrons and muons with momenta in the range $1.0 < p_{\ell} < 3.0\,\gevc$, thus bridging the gap between the low-momentum two-photon-fusion samples and the high-momentum $\epem \to \ell^+ \ell^- (\gamma)$ samples.  We select two tracks with $|dz| < 5$ cm, and $|dr| < 2$ cm. The invariant mass of the reconstructed $J/\psi$ candidate is required to be within the range $2.8 < M_{\ell^+\ell^-} < 3.3\,\gevcc$ and a vertex fit of the two tracks is performed, retaining all candidates with a successful fit. To reduce the contamination from low-multiplicity events that contain leptons not originating from a $J/\psi$ decay, the ratio between the (event-based) second-order and zeroth-order Fox-Wolfram moment~\cite{Fox:1978vu}, $R_2$, is required to be below 0.4. This requirement allows for a measurement of the efficiency in a hadron-enriched environment. To further clean up residual contamination of low multiplicity QED events, a minimal requirement of five tracks per event is enforced, and to ensure better identification of electrons, both candidates are required to have a cluster in the calorimeter matched to their track through the inner detectors. For this sample, we employ the peak-fitting technique, but also utilize tag-and-probe to increase the signal-to-noise ratio, applying a cut at 0.95 on the global likelihood ratio (see Sec.~\ref{sec:probabilities}) for the electron (muon) hypothesis of the tag track.
Figure~\ref{fig:Jpsi_masses} shows the $J/\psi$ peaks and the level of signal-to-background that is achieved with our selection.

\bigskip

\begin{figure*}[htbp]
    \begin{center}
        \includegraphics[width=0.8\textwidth]{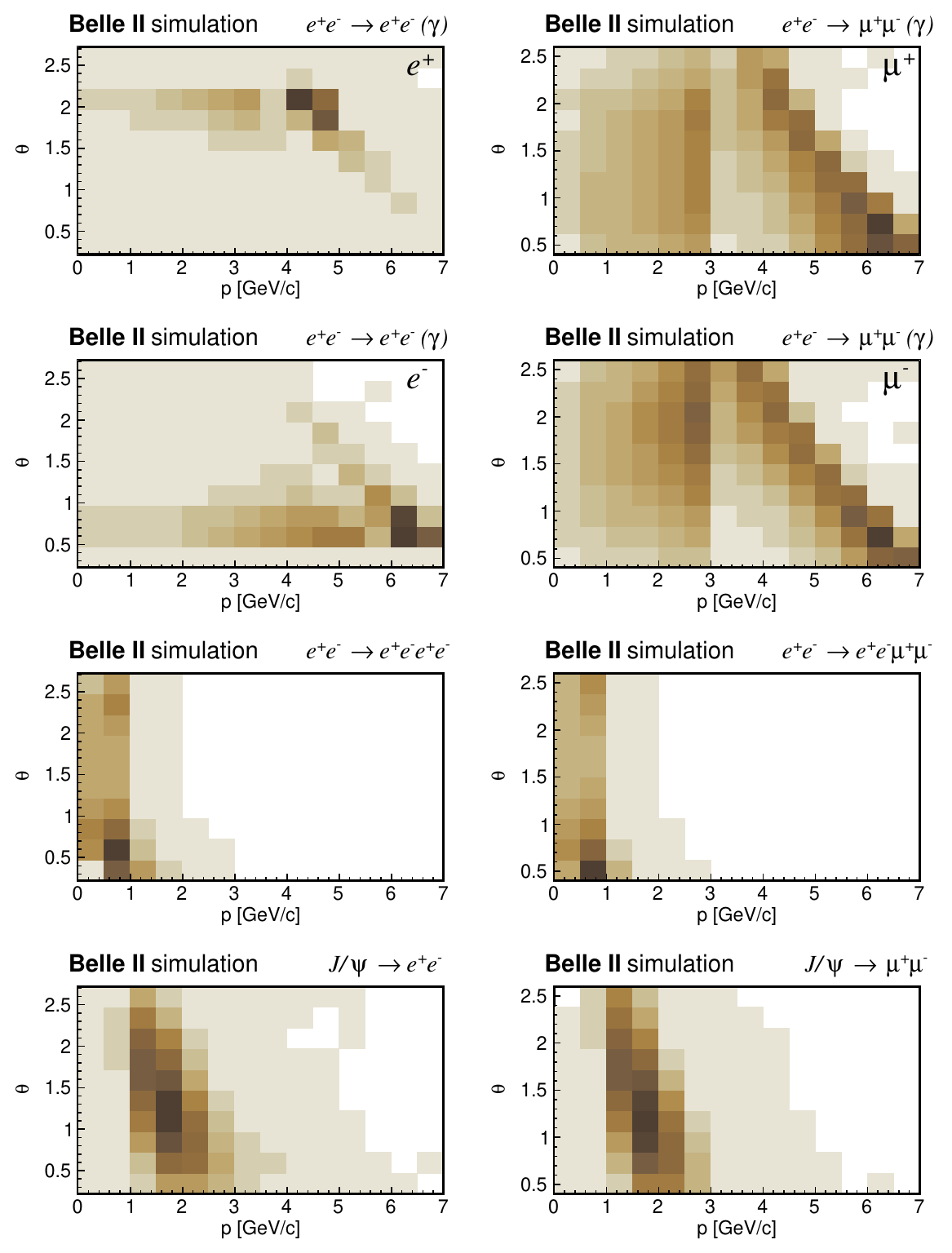}
        \caption{Density (in linear scale) of the polar angle and momentum distributions obtained from the simulation for all the electron (left) and muon (right) samples considered. From top to bottom: $\epem \to \ell^+\ell^- (\gamma)$ (positive tracks only), $\epem \to \ell^+\ell^- (\gamma)$ (negative tracks only), $\epem \to \epem \ell^+\ell^-$, and $J/\psi \to \ell^+\ell^-$.}
        \label{fig:Signal_shapes}
    \end{center}
\end{figure*}

Figure~\ref{fig:Signal_shapes} shows the two-dimensional distributions of polar angle and momentum, obtained from the simulation, of all the lepton samples described above.

\subsection{Hadron samples}
\label{subsec:hIDsamples}

To measure the probability that a hadron ($\pi$, $K$, $p$) passes a certain PID selection meant to select leptons, we use samples in which the parent particle decays to two charged particles, to which no PID criteria are applied. We select $K^0 \to \pi^+\pi^-$, $D^{*+} \to D^0 \pi^+_{\rm soft}$ with $D^0 \to K^- \pi^+$, and $\Lambda \to p \pi^-$ \footnote{Unless explicitly stated otherwise, charge conjugation is always implied throughout this paper.} samples. The context indicates the species of the particles involved: a $K^0$ decaying to two charged particles almost always decays to pions, in the $D^{*+}$ decay chain described above, the charge of the (soft) pion $\pi^+_{\rm soft}$ from the $D^{*+}$ decay indicates which of the two daughters of the $D^0$ is the kaon (the track with charge opposite to that of the soft pion), and the kinematics of the $\Lambda$ decay are such that the particle carrying most of the $\Lambda$ momentum is the proton. For more details about the selection of these samples, we refer to Ref.~\cite{Belle-II:2025tpe}.

\section{Local likelihoods}
\label{sec:likelihoods}

All \BelleII subdetectors, with the exception of the PXD, contribute to PID. Subdetector $d$, performing the set of measurements $\vec{x}_d$ on a track of momentum $\vec{p}$, provides the local likelihood $\mathcal{L}^d_{\alpha}(\vec{x}_d | \vec{p})$ for the  particle hypothesis $\alpha$: $\alpha \in \{e$, $\mu$, $\pi$, $K$, $p$, $d\}$. In this section, we give a brief overview of how each subdetector determines $\mathcal{L}^d_{\alpha}$, referring to Ref.~\cite{Belle-II:2025tpe} for further details.

The SVD measures the specific energy loss (that is, the energy loss per unit distance) due to ionization \dEdx* \cite{Chapter34_pdg} typically eight times for a particle traversing all its four double-sided strip layers. In order to reduce the bias from the long tail of the Landau distribution, the two highest measurements are excluded from the average. Also the CDC distinguishes between different particle species by measuring \dEdx* in each of its drift cells. For a track fully traversing the subdetector, typically 40--60 individual measurements are obtained, of which the lowest 5\% and highest 25\% are discarded to avoid biases.

The TOP measures the arrival time of Cherenkov photons that are produced by the passage of the charged particle through one of its quartz bars. For a given momentum, and point and angle of impact, analytical probability density functions are calculated for the expected density of photons in each readout channel for each particle hypothesis; the Poisson probability of detecting $N$ photons for a given track is also taken into account~\cite{Staric:2008}. The readout electronics of the ARICH photon detectors only provides binary signals: whether a pixel fired or not. For a given track and for each particle hypothesis, the expected number of photons to hit each pixel is computed and the local likelihood is calculated.

The ECL measures the energy deposited by a particle in its crystals. The ratio of its deposited energy to its momentum strongly depends on the particle species. The species-dependent likelihood for this ratio is calculated using Gaussian kernel density-estimation templates~\cite{Cranmer:2000du} determined from simulated data of isotropically distributed particles. They are determined separately for each region of the Cartesian product of three $\theta$ regions, three $p$ regions, and the two polarities of electric charge, $q$. The boundaries of the $\theta$ regions are $17^{\circ}$, $32^{\circ}$, $128^{\circ}$, and $150^{\circ}$, which are the boundaries of the ECL’s forward end cap, barrel, and backward end cap. The boundaries of the two lower $p$ regions are $0.2\,\gevc$, $0.6\,\gevc$, and $1.0\,\gevc$; the third region contains all momenta greater than $1.0\,\gevc$.

Muons with momenta greater than $1\,\gevc$ typically traverse the whole KLM; hadrons typically stop in its first layers; and electrons rarely reach it. So we use the depth of hits in the KLM, along with information about their lateral shape, to distinguish between particle species. We extrapolate each track from the CDC into the KLM using a Kalman filter, assuming it is a stable muon and accounting for multiple scattering and ionization energy loss. The species-dependent likelihood is the product of the likelihoods for the observed depth and the lateral shape, neglecting the potential correlation between the two.

\section{PID probabilities}
\label{sec:probabilities}

Assuming the local likelihoods are well-formed and uncorrelated, we multiply them to form global species-dependent likelihoods
\begin{equation}
\mathcal{L}_\alpha \equiv \prod_d \mathcal{L}^d_\alpha.
\label{eqn:basic_pid_prob}
\end{equation}
Using Bayes' theorem and the law of total probability, and assuming a uniform prior probability for all species, 
we compute the likelihood ratio
\begin{equation}
    P_{\alpha} = \frac{\mathcal{L}_\alpha}{\sum_\gamma \mathcal{L}_\gamma},
    \label{eqn:unweighted_pid_prob}
\end{equation}
where the sum in the denominator runs over all species~($e$, $\mu$, $\pi$, $K$, $p$, $d$).
In the special case where all considered particle species are present with the same abundances, this likelihood ratio corresponds to the probability that the particle under study is of species $\alpha$.
In many analyses, we only need to decide between a subset of possible species and restrict the values of considered species $\gamma$.
The binary likelihood ratio, restricting $\gamma$ to two species,
\begin{equation}
    P_{\alpha/\beta} \equiv \frac{\mathcal{L}_\alpha}{\mathcal{L}_\alpha + \mathcal{L}_\beta},
    \label{eqn:unweighted_bin_pid_prob}
\end{equation}
is especially useful. 
Throughout the rest of the paper we will only show the performance of the global likelihood ratios, but binary ratios are supported and utilized in physics analyses.

Most particles do not enter all subdetectors.
It is very rare that a particle enters both the TOP and the ARICH, and no particles with momenta below \SI{500}{MeV/c} reach the KLM.
When a particle does not enter a subdetector, we assign $\mathcal{L}^d_\alpha = 1$ for all $\alpha$.

These probabilities are easy to implement and maintain in \basfii. For example, we can easily exclude local likelihoods from subdetectors that were not well calibrated in some period. However, these probabilities have significant disadvantages that limit performance.
They neglect correlations, though we expect them, since the likelihoods all depend on the same track parameters.
They also do not account for inefficient or uninformative detectors, allowing probabilities to possibly favor one species over another based mostly on statistical fluctuations.
The probabilities also assume that the local likelihoods are well formed and normalized such that no species is favored over another.

As we discussed in Ref.~\cite{Belle-II:2025tpe}, the approach of simple probabilities gives suboptimal performance at \BelleII and significant improvement can be obtained by either reweighting the individual subdetectors' likelihoods in order to exploit the advantages of each subdetector in the different regions of the phase space or by moving to a multivariate machine-learning based approach.

\subsection{Boosted decision tree-based lepton ID}
\label{subsec:BDTs}

At low momentum ($p \leq 1\,\gevc$) the $E/p$ ratio does not provide sufficient discrimination power for effective PID. On the other hand, the calorimeter provides a number of variables (ratios of energy deposited in the inner/outer crystals of a cluster, lateral moment, Zernike moments, etc.) that characterize ECL clusters and can be exploited to enhance the discrimination power.
This is done using a boosted decision tree (BDT) classifier that combines these additional variables with the other subdetectors’ (with the exclusion of the SVD) likelihood ratios. 

Algorithms are trained both for binary $e/\pi$ and $\mu/\pi$ separation, as well as in global mode to separate a lepton hypothesis from all other hypotheses. Thus, they directly compare to the binary (global) likelihood ratio of Eq.~\ref{eqn:unweighted_bin_pid_prob} (Eq.~\ref{eqn:unweighted_pid_prob}), respectively.
The ECL observables’ distributions generally depend on the particle’s momentum and charge, as well as on geometrical effects related to the calorimeter structure. Furthermore, the likelihood ratios of the other subdetectors included in the BDT are often defined only in specific subsets of the full detector acceptance. Therefore, a categorization is performed by training independent classifiers in each of the 18 $(p, \theta, q)$ categories defined for ECL PID in Sec.~\ref{sec:likelihoods}. 
An additional novel variable, which consists of a multivariate classifier based on characteristics of the waveform pulse shape read out from the ECL crystals to separate electromagnetic and hadronic-interacting particles~\cite{Longo:2020zqt}, here referred to as “PSD MVA”, has been included. This variable improves separation for electrons against hadrons. More details about the input variables and the training of the BDT are provided in Ref.~\cite{Milesi:2020esq}.

We note that, unlike the simple likelihood-based probabilities, the BDT algorithm optimally selects a sequence of node-splitting criteria based on the input variables, ultimately weighting the importance of each variable in the combination. As a result, the inclusion of subdetectors that perform poorly in specific regions of phase space does not deteriorate the overall performance in those problematic regions, while it generally improves the discrimination power elsewhere.

\section{PID efficiencies and mis-ID rates}
\label{sec:eff_and_misID}

We use the control samples described in Sec.~\ref{sec:data} to measure the efficiency, which is the probability that a particle of species $\alpha$ passes the $\alpha$-ID selection, and the $\beta$-as-$\alpha$ mis-identification (mis-ID) rate, which is the probability that a particle of species $\beta$ meets the $\alpha$-ID requirements. We measure efficiencies and mis-ID rates both in data and simulation, in subregions of momentum and polar angle of the candidate tracks and separately for positively and negatively charged particles. The difference in PID performance between simulation and data is then used in the physics analysis to correct the simulation and make it more similar to the performance observed in the data.

At \BelleII we measure efficiencies and mis-ID rates using two different procedures, one based on the tag-and-probe technique and the other employing the fitting of an invariant mass distribution. The first is utilized in samples in which no peaking structures are present ($\epem \to \epem (\gamma), \; \mu^+\mu^-\; \gamma, \epem \ell^+\ell^-$), the second is used for the $J/\psi \to \ell^+\ell^-$  samples and for the samples providing hadrons ($D^0 \to K^- \pi^+$, $\Lambda \to p \pi^-$, $K^0 \to \pi^+\pi^-$). 

With the tag-and-probe technique, the $\alpha$-ID efficiency $\varepsilon_{\alpha}$ is measured as
\begin{equation}
    \varepsilon_{\alpha} = \frac{N^{\rm probe} - B^{\rm probe}}{N^{\rm tag} - B^{\rm tag}} \; ,
\end{equation}
where $N^{\rm tag}$ is the number of data events passing the selection for a specific control channel that provides particles of species $\alpha$, including the PID selection on the designated tag track, and $N^{\rm probe}$ is the number of events passing the same selection criteria and the additional $\alpha$-ID requirement on the designated probe track. $B^{\rm tag}$ and $B^{\rm probe}$ are the number of background events that satisfy the same criteria imposed to obtain $N^{\rm tag}$ and $N^{\rm probe}$, respectively. In all cases we derive the number of background events from the simulation, rescaling it to the equivalent luminosity of the data. If both lepton tracks in the event pass the requirements for the tag, both candidates are used as probes to measure PID efficiencies. To account for variations of performance across the phase space, the procedure can be applied to small subregions of momentum and polar angle. The tag-and-probe technique can be utilized in the same way to measure the $\alpha$-as-$\beta$ mis-ID rate using a sample providing particles of species $\alpha$ and applying $\beta$-ID criteria to the probe track. 
For the systematic uncertainties, we vary the normalization of each background component by a factor that conservatively covers the uncertainty associated with each background mode. For the $\epem \to \epem \ell^+\ell^-$ samples, we also include the uncertainties on the scaling factors that are used to correct the shape of the simulated backgrounds and the uncertainties on the hadron-to-lepton mis-ID rates that are applied to the simulation. For the $\epem \to \epem \, (\gamma)$ sample, we also compute a systematic uncertainty related to the trigger by taking the difference between the efficiencies evaluated with the nominal method and those obtained without applying any trigger selection.

The peak-fitting technique is based on estimating the number of correctly reconstructed decays of the parent particle before and after PID selection criteria are applied to one of the daughter tracks. The efficiency is computed as the ratio between the signal yield obtained when the $\alpha$-ID criteria are applied to the particle of species $\alpha$ and the yield when this PID selection is removed. 

For the $J/\psi$ sample, in each separate subregion of momentum and polar angle (and separately for positive and negative charges), we distinguish between correctly reconstructed $J/\psi$ events and random combinations of tracks by fitting the $m(\ell^+\ell^-)$ distributions with the sum of a Gaussian, a bifurcated Gaussian, and a Crystal Ball~\cite{Gaiser:Phd, Skwarnicki:1986xj} function for the signal component, and a second-order Chebyshev polynomial for the background. The statistical uncertainty on the fitted yields is propagated to the efficiency, and systematic contributions are evaluated by varying the parameters that are kept fixed in the fit by their uncertainties, determined from a fit to the $m(\ell^+\ell^-)$ distribution in the whole phase space.

For the hadron samples we fit the invariant mass distribution obtained before applying any PID selection criteria, and determine the sWeights~\cite{Pivk:2004ty} for each event. This procedure is based on the assumption (which we verify) that the sWeights are uncorrelated with the quantities on which the PID discrimination is based and can thus be reliably utilized to subtract the background under the peaks we are considering. The $\alpha$-ID efficiency is computed as the sum of the sWeights for events in which the relevant track passes the $\alpha$-ID criteria divided by the sum of the sWeights of all events considered. Analogously, the $\beta$-as-$\alpha$ mis-ID rate is determined on a sample providing particles of species $\beta$ and taking the ratio of the sum of the sWeights of the events passing the $\alpha$-ID criteria and the sum of the sWeights of all events considered. For the systematic uncertainties, we include two sources: the first is the difference in the results obtained when using two alternative probability density functions for the background component, while the second is estimated on the simulation by taking the difference between the efficiency obtained with the simulation-matched assignment of particles and the one obtained with the nominal sWeights procedure.
Details on the probability density functions used for the hadron samples and on the software infrastructure can be found in Ref.~\cite{Belle-II:2025tpe}.

\section{Lepton-identification performance}
\label{sec:performance}

We use the techniques illustrated in the previous section to characterize the PID performance of \BelleII in terms of lepton identification efficiency and hadron-as-lepton mis-ID rates ($e$-as-$\mu$ and $\mu$-as-$e$ mis-ID rates are at or below the permille level). In case the lepton under study emits a photon while traversing the beam pipe or the innermost layers of the tracking devices (this happens quite frequently for the electrons), we do not make any attempt to recover the four-momentum of the lepton before the interaction with the detector material, thus $p$ and $\theta$ will always refer to the reconstructed track.

Both for electron and muon ID, the most problematic region of the phase space is the low momentum ($ p \lesssim 1\,\gevc$) region, where particles often do not traverse the ECL and KLM subdetectors, so PID mostly relies on the \dEdx* measurements provided by the tracking devices. At higher momenta, and especially in the barrel region, the lepton/hadron separation capabilities become much stronger, particularly for the electron case.

\begin{figure}[htbp]
    \begin{center}
        \includegraphics[width=0.9\textwidth]{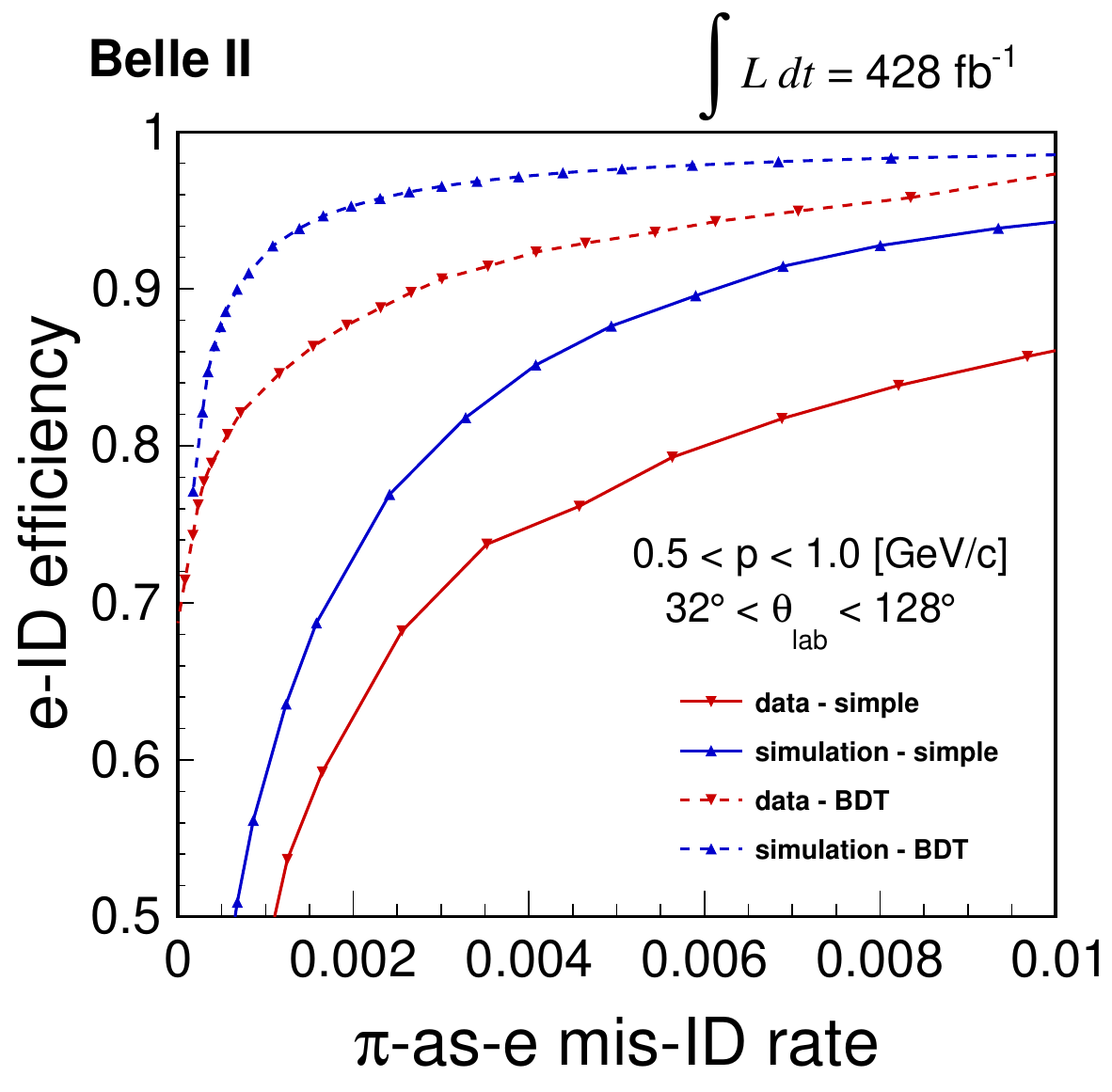}
        \caption{Electron identification efficiency as a function of the $\pi$-as-$e$ mis-ID rate, measured on the $\epem \, e^+e^-$ and $D^*$ control samples, for the low momentum region ($0.5 < p < 1.0\,\gevc$) and restricted to the ECL barrel region ($32^{\circ} < \theta < 128^{\circ}$). Red downward (blue upward) pointing triangles represent the data (simulation), while solid (dashed) lines correspond to the simple (BDT based) probabilities.} 
        \label{fig:eID_roc}
    \end{center}
\end{figure}

\begin{figure}[htbp]
    \begin{center}
        \includegraphics[width=0.9\textwidth]{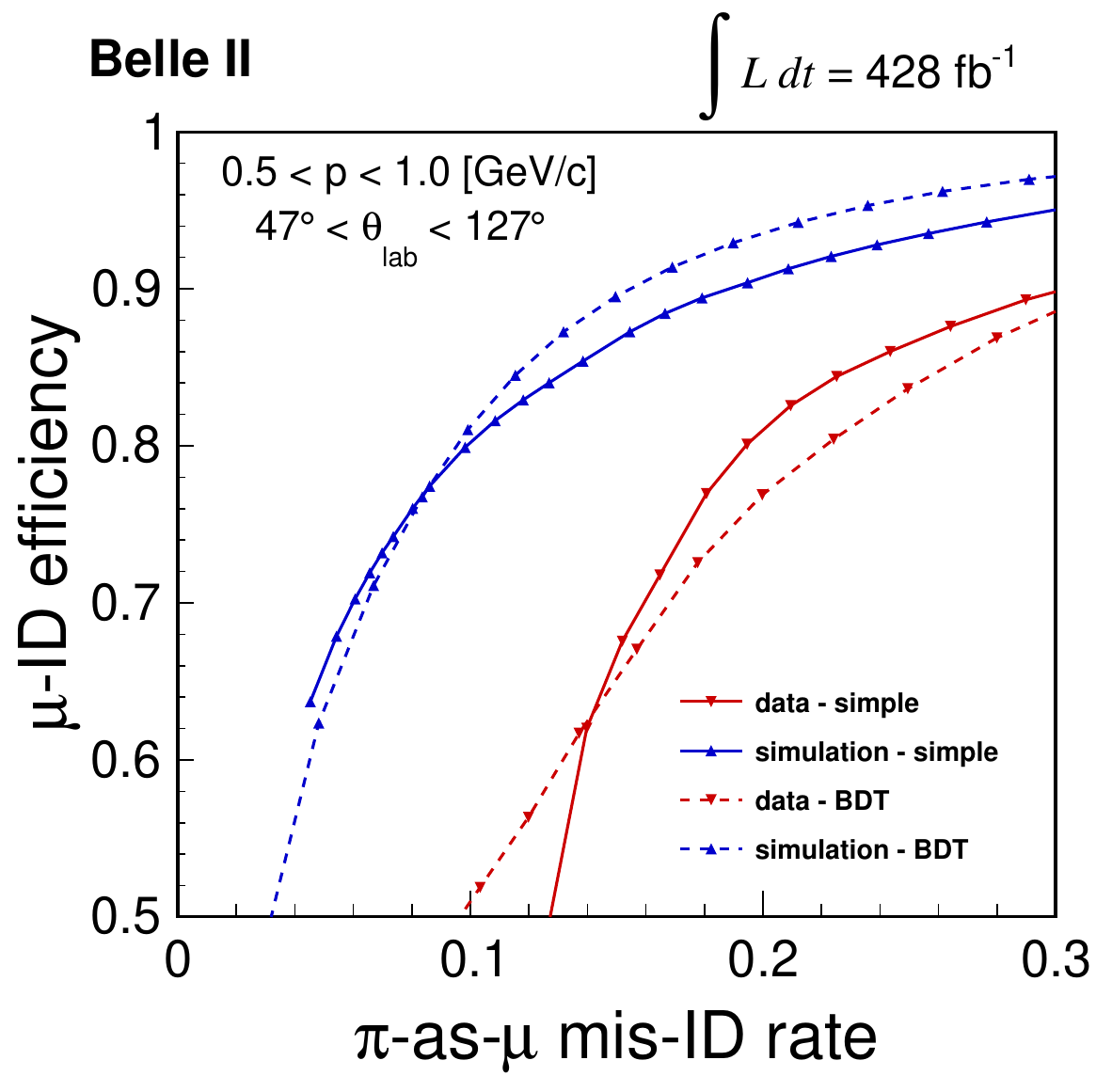}
        \caption{Muon identification efficiency as a function of the $\pi$-as-$\mu$ mis-ID rate, measured on the $\epem \, \mu^+\mu^-$ and $D^*$ control samples, for the low momentum region ($0.5 < p < 1.0\,\gevc$) and restricted to the KLM barrel region ($47^{\circ} < \theta < 127^{\circ}$). Red downward (blue upward) pointing triangles represent the data (simulation), while solid (dashed) lines correspond to the simple (BDT based) probabilities.} 
        \label{fig:muID_roc}
    \end{center}
\end{figure}

In Figs.~\ref{fig:eID_roc} and \ref{fig:muID_roc} we show the lepton vs pion 
discrimination power, measured for the low-momentum particles that fall in the barrel region.

For the electrons, we denote the global likelihood computed with the exclusion of the SVD and TOP subdetectors as \emph{simple probability}.\footnote{We exclude the SVD and TOP as they are shown to degrade the performance, due to imperfect calibrations of the SVD \dEdx*, and to the fact that modeling of showering effects makes it difficult to identify electrons with the TOP.} Its performance is compared to that of the (global) BDT-based selection. Overall the performance for electron identification is very good with electron ID efficiencies between 50\% and 98\% for $\pi$-as-$e$ mis-ID rates between 0.1\% and 1\%. It is apparent that the BDT-based algorithm has significantly better performance than the simple probability, thanks to the use of more information from the ECL and to the better handling of the correlations among different variables. For this reason, the BDT-based selection is the strategy typically used in physics analyses, even though the simple approach can still achieve excellent $e$/$\pi$ separation. It is also noticeable that the performance in data is significantly worse than what is expected in the simulation. One of the main reasons for this discrepancy (that affects also the muon ID) is the sensitivity of the CDC readout to the beam-injection backgrounds. This effect (which is not reproduced by the simulation) is extensively discussed in Ref.~\cite{Belle-II:2025tpe} and results in a bias on the \dEdx* measurement during the few milliseconds that follow a beam $e^{\pm}$ bunch injection.

In the case of muons, for momenta below $1\,\gevc$, we can obtain efficiencies above 80\% only by accepting $\pi$-as-$\mu$ mis-ID rates above the 18 percent level. We observe that in data the simple probability (obtained by combining the likelihoods of all subdetectors with the exception of the SVD) obtains in general better performance compared to the BDT-based selection (whereas the opposite is true in simulation). The better performance of the simple probability in data is established also at $p > 1\,\gevc$, so this quantity is recommended for use in physics analysis. 


\subsection{Lepton-ID efficiencies}

We evaluate the electron ID and muon ID efficiencies as a function of momentum and polar angle for a fixed cut on the simple or BDT-based probabilities in the control samples described in Sec.~\ref{sec:data}. 

\begin{figure*}[htbp]
    \begin{center}
        \begin{tabular}{c}
            \includegraphics[width=0.67\textwidth]{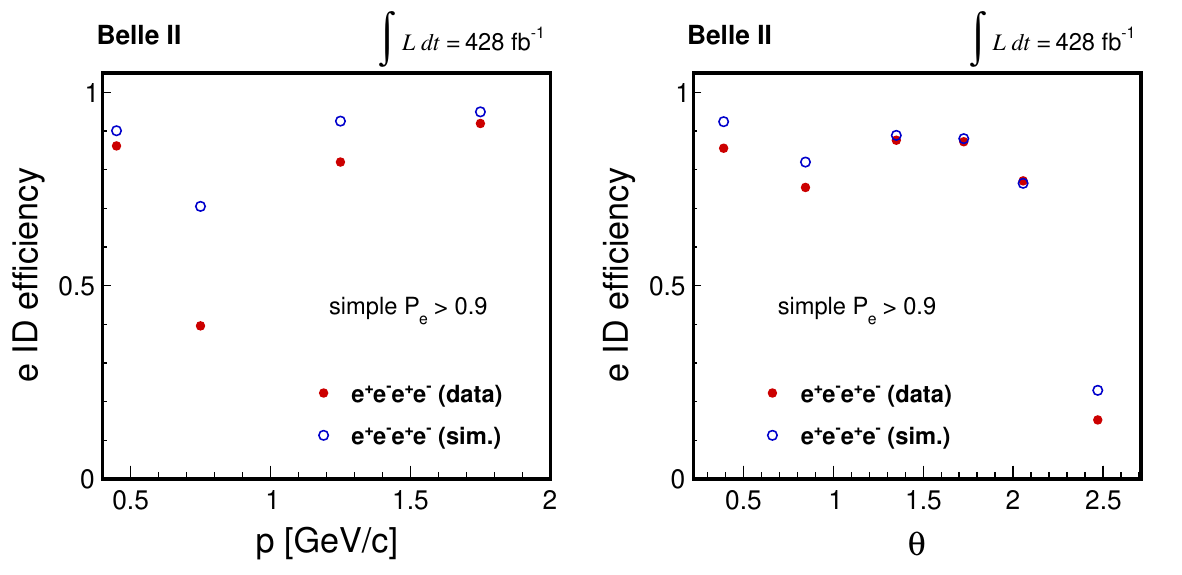} \\
            \includegraphics[width=0.67\textwidth]{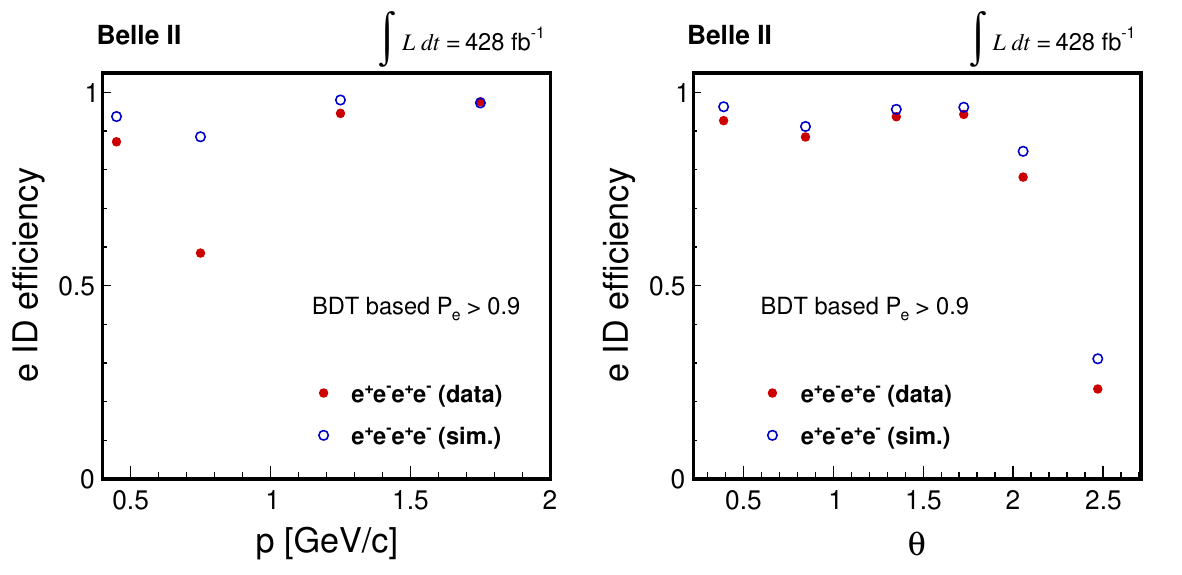} 
        \end{tabular}
        \caption{Electron ID efficiency as a function of the momentum (left) and polar angle (right) as measured in the $\epem \, \epem$ sample using a simple probability $P_e > 0.9$ selection (top) and a BDT-based probability $P_e > 0.9$ selection (bottom). Red filled (blue empty) markers represent the data (simulation).} 
        \label{fig:eID_eell}
    \end{center}
\end{figure*}

Figure~\ref{fig:eID_eell} shows the electron ID performance as measured in the $\epem \, \epem$ sample for the simple $P_e > 0.9$ selection and BDT-based $P_e > 0.9$. Overall the simulation reproduces reasonably well the data, with the notable exception of the momentum region around $p \sim 0.7\,\gevc$, where the impact of the injection backgrounds affecting the CDC \dEdx* measurement is largest. We also see a large drop of performance at $\theta \gtrsim 2$ (backward region), where PID mostly relies on the \dEdx* measurement of the CDC.

\begin{figure*}[htbp]
    \begin{center}
        \begin{tabular}{c}
            \includegraphics[width=0.67\textwidth]{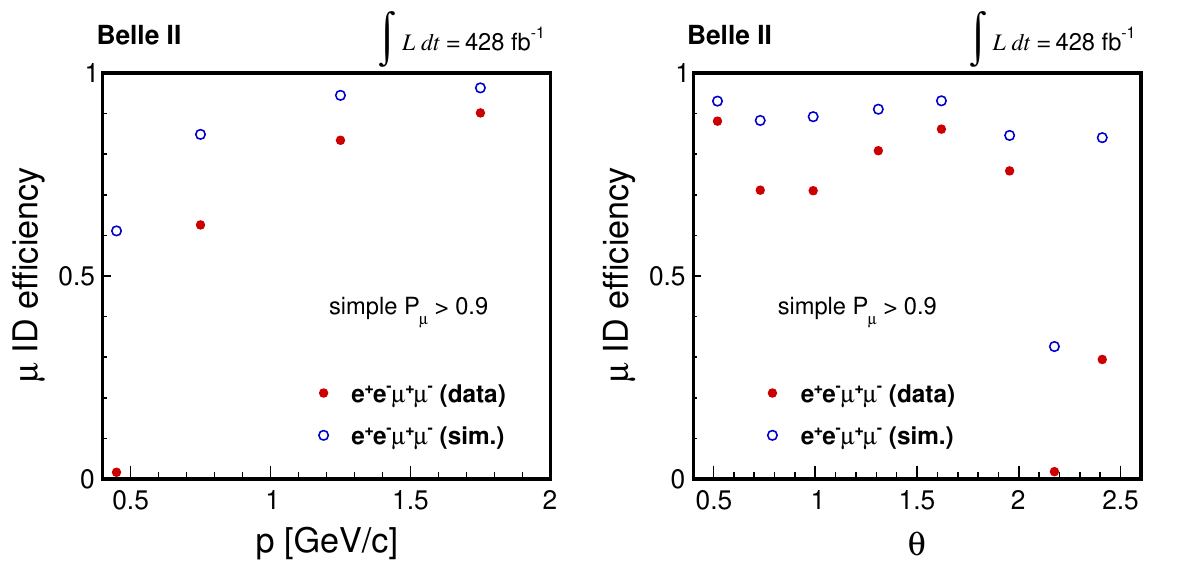} \\
            \includegraphics[width=0.67\textwidth]{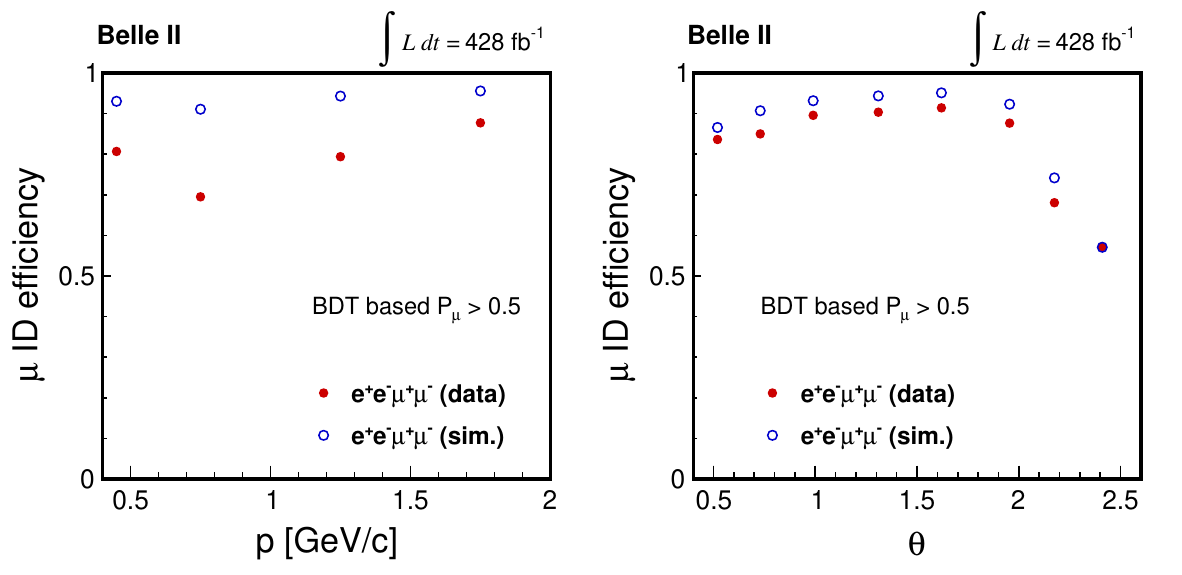} 
        \end{tabular}
        \caption{Muon ID efficiency as a function of the momentum (left) and polar angle (right) as measured in the $\epem \, \mu^+\mu^-$ sample using a simple probability $P_{\mu} > 0.9$ selection (top) and a BDT-based probability $P_{\mu} > 0.5$ selection (bottom). Red filled (blue empty) markers represent the data (simulation).} 
        \label{fig:muID_eell}
    \end{center}
\end{figure*}

In Fig.~\ref{fig:muID_eell} we show the analogous performance plots for muon ID for the simple $P_{\mu} > 0.9$ selection and BDT-based $P_{\mu} > 0.5$ (we choose the latter working point as it gives more similar performance to the simple probability selection). At very low momentum and in the backward region, the simple probability mostly relies on the \dEdx* measurement of the CDC, which has low discrimination power due to the similarity of the masses of muons and pions and the sensitivity to injection backgrounds described above. On the other hand the BDT-based probability, which relies also on additional information from the ECL, shows better performance in this corner of the phase space and a more uniform behavior as a function of the polar angle.


\begin{figure*}[htbp]
    \begin{center}
        \begin{tabular}{c}
            \includegraphics[width=0.67\textwidth]{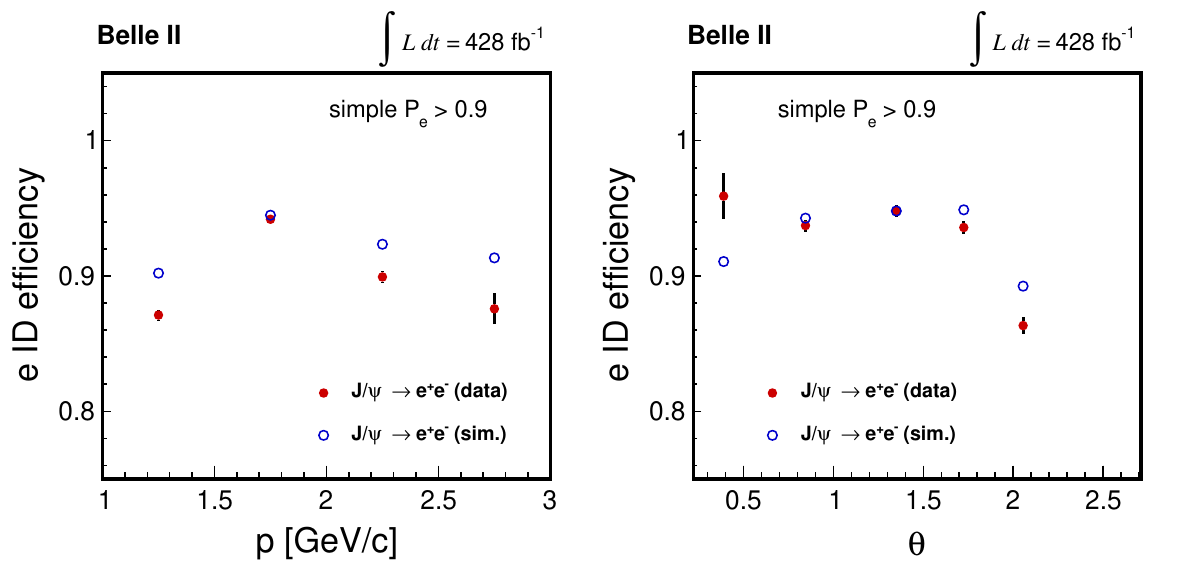} \\
            \includegraphics[width=0.67\textwidth]{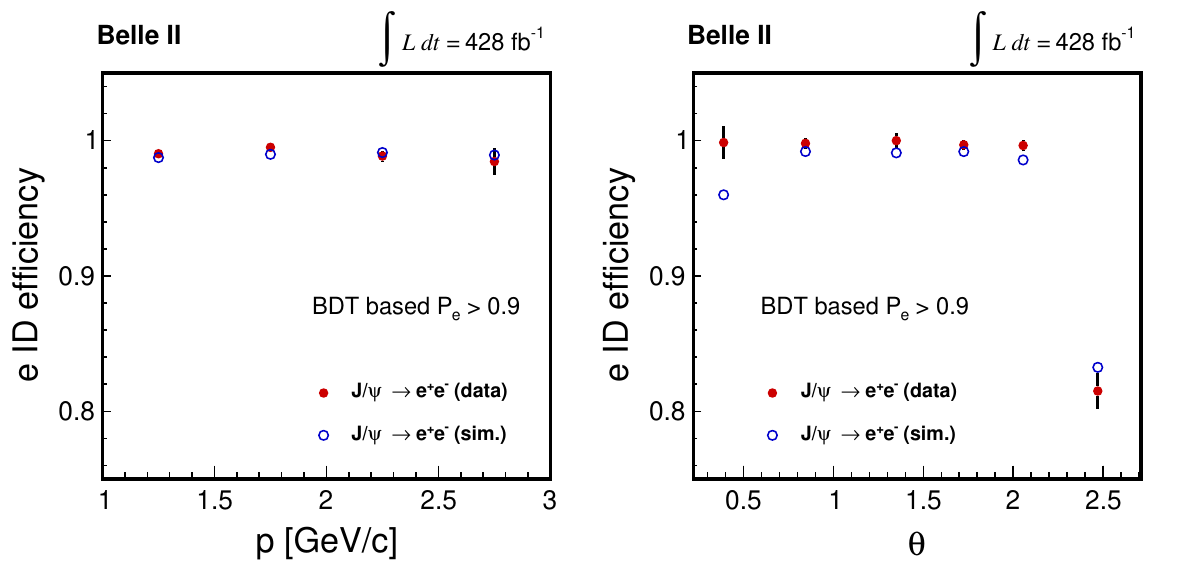} 
        \end{tabular}
        \caption{Electron ID efficiency as a function of the momentum (left) and polar angle (right) as measured in the $J/\psi \to \epem$ sample using a simple probability $P_e > 0.9$ selection (top) and a BDT-based probability $P_e > 0.9$ selection (bottom). Red filled (blue empty) markers represent the data (simulation).} 
        \label{fig:eID_jpsill}
    \end{center}
\end{figure*}

The intermediate momentum region ($1.0 < p < 3.0\,\gevc$) is studied with the $J/\psi$ samples. For the electron ID (Fig.~\ref{fig:eID_jpsill}) the performance is quite stable as a function of the momentum, while we see a large drop of performance in the very backward region. The BDT-based selection confirms its performance advantage over the likelihood-based simple probability.

\begin{figure*}[htbp]
    \begin{center}
        \begin{tabular}{c}
            \includegraphics[width=0.67\textwidth]{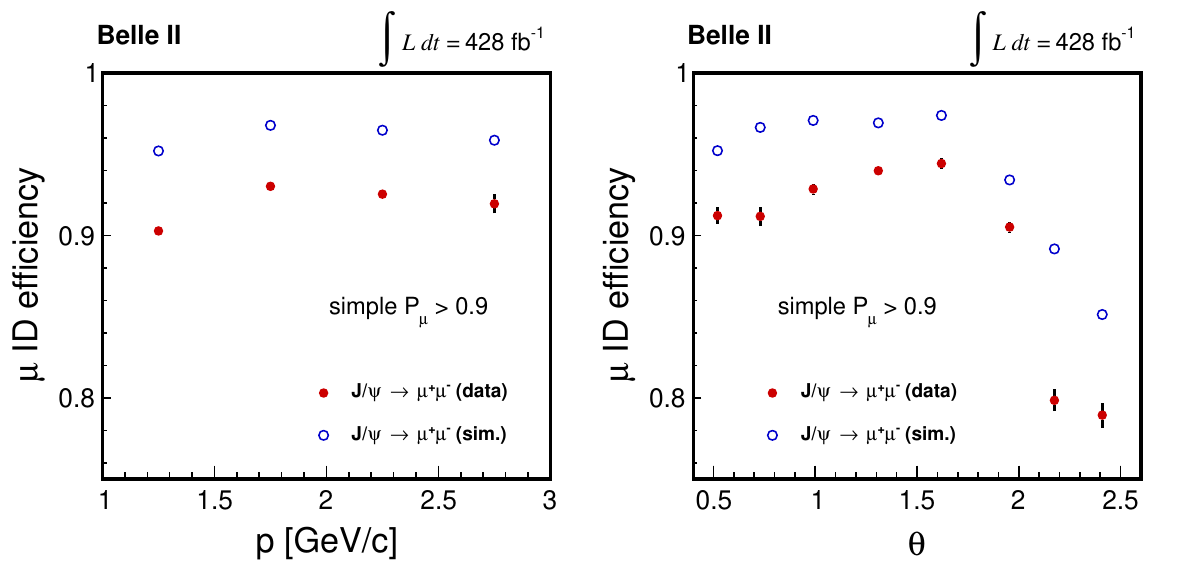} \\
            \includegraphics[width=0.67\textwidth]{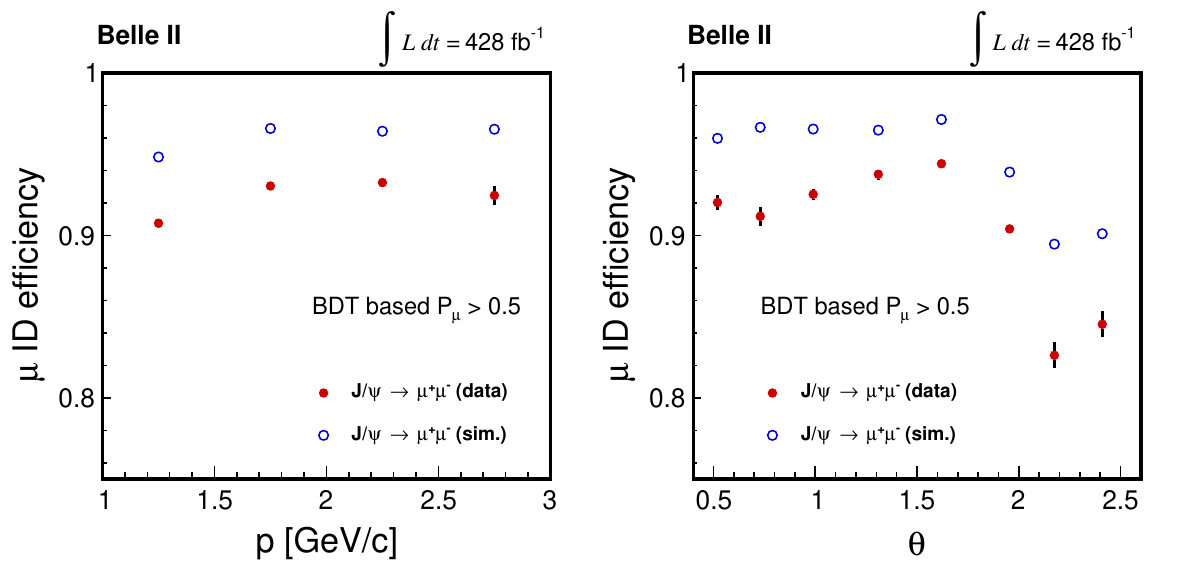} 
        \end{tabular}
        \caption{Muon ID efficiency as a function of the momentum (left) and polar angle (right) as measured in the $J/\psi \to \mu^+\mu^-$ sample using a simple probability $P_{\mu} > 0.9$ selection (top) and a BDT-based probability $P_{\mu} > 0.5$ selection (bottom). Red filled (blue empty) markers represent the data (simulation).} 
        \label{fig:muID_jpsill}
    \end{center}
\end{figure*}

Also for the muons (Fig.~\ref{fig:muID_jpsill}), the performance is stable in this momentum range, with the simple and BDT-based selections giving similar results in terms of efficiency.

\begin{figure*}[htbp]
    \begin{center}
        \begin{tabular}{c}
            \includegraphics[width=0.67\textwidth]{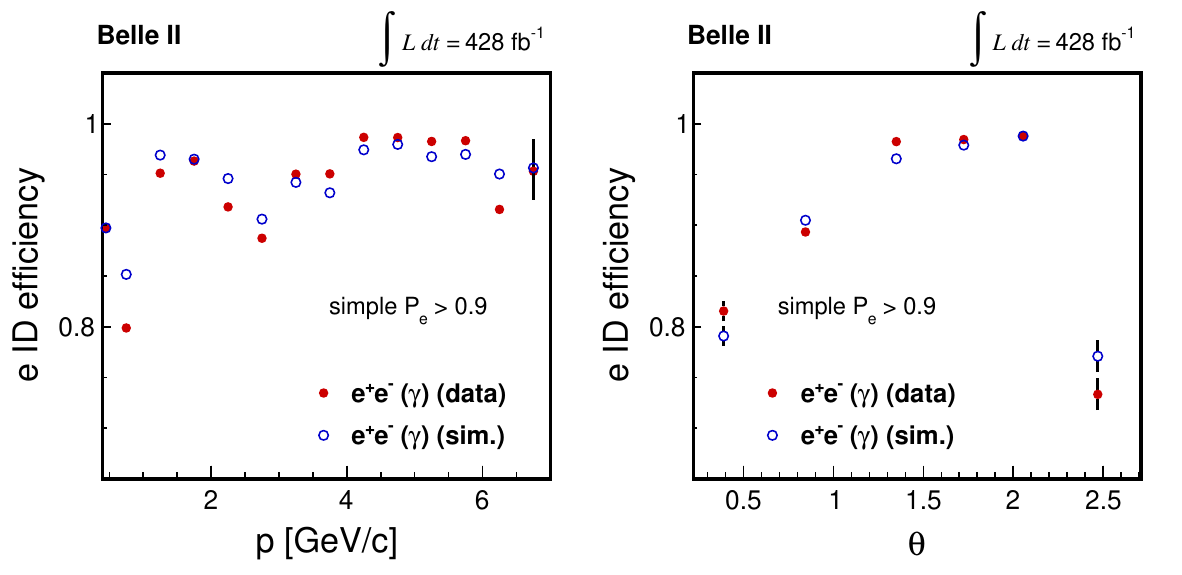} \\
            \includegraphics[width=0.67\textwidth]{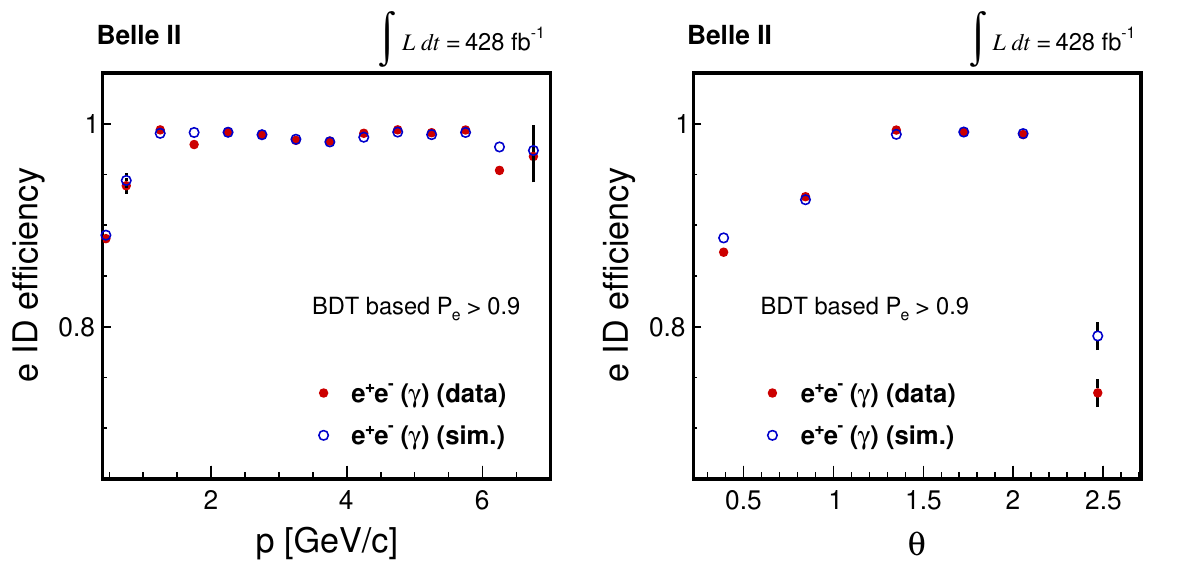} 
        \end{tabular}
        \caption{Electron ID efficiency as a function of the momentum (left) and polar angle (right) as measured in the $\epem \to \epem \, (\gamma)$ sample using a simple probability $P_e > 0.9$ selection (top) and a BDT-based probability $P_e > 0.9$ selection (bottom). Red filled (blue empty) markers represent the data (simulation).} 
        \label{fig:eID_llg}
    \end{center}
\end{figure*}

\begin{figure*}[htbp]
    \begin{center}
        \begin{tabular}{c}
            \includegraphics[width=0.67\textwidth]{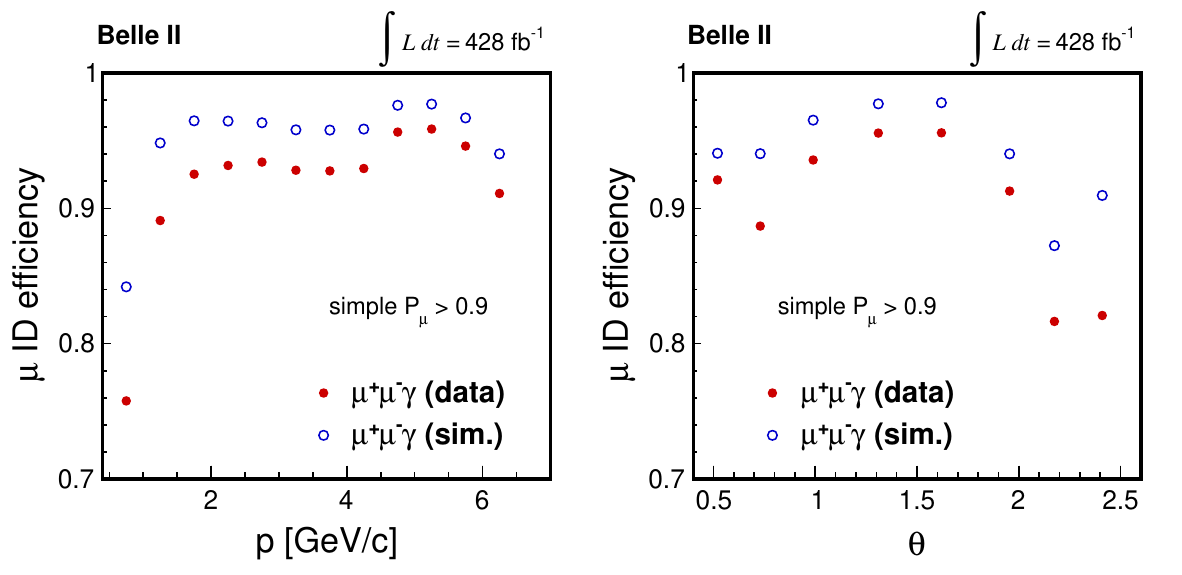} \\
            \includegraphics[width=0.67\textwidth]{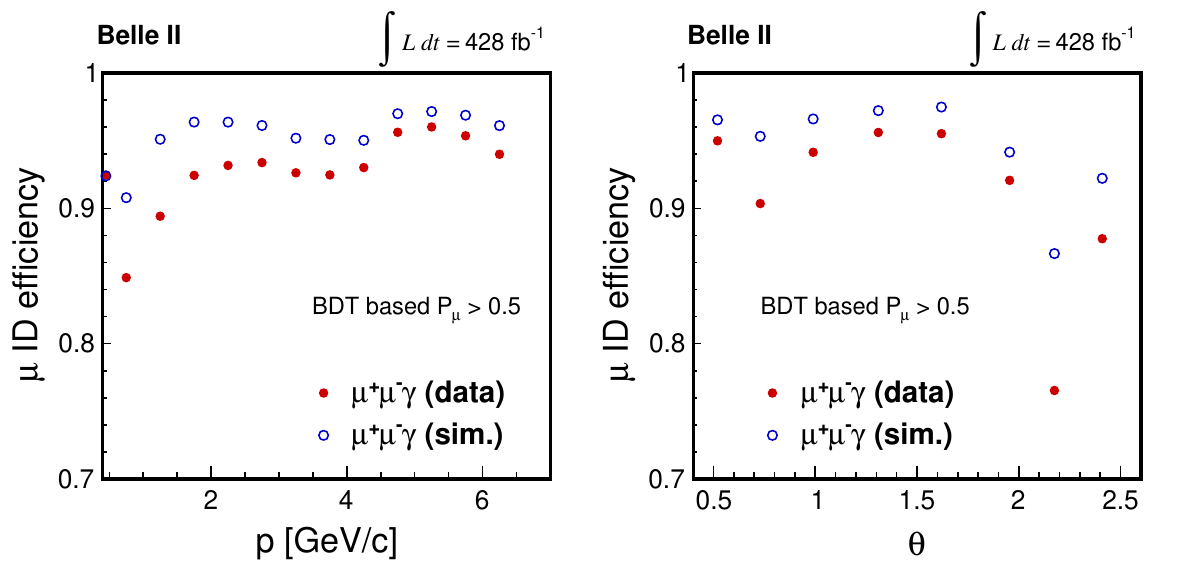} 
        \end{tabular}
        \caption{Muon ID efficiency as a function of the momentum (left) and polar angle (right) as measured in the $\epem \to \mu^+\mu^- \, \gamma$ sample using a simple probability $P_{\mu} > 0.9$ selection (top) and a BDT-based probability $P_{\mu} > 0.5$ selection (bottom). Red filled (blue empty) markers represent the data (simulation).} 
        \label{fig:muID_llg}
    \end{center}
\end{figure*}

The $\epem \to \epem \, (\gamma)$ and $\epem \to \mu^+\mu^- \, \gamma$ samples cover a wide momentum range, which overlaps with the samples discussed above and extends to the kinematic limit attainable at our experiment. 
Figures~\ref{fig:eID_llg} and \ref{fig:muID_llg} show the electron ID and muon ID performance for the simple and BDT-based selections. The step that is visible in the muon ID performance above $4\,\gevc$ is mostly due to the correlation between momentum and polar angle (every bin in the momentum plot averages over the whole $\theta$ range) and the initial ($p$, $\theta$) distribution of the control sample (see Fig.~\ref{fig:Signal_shapes}).

\subsection{Hadron-as-lepton mis-ID rates}

\begin{figure*}[htbp]
    \begin{center}
        \includegraphics[width=0.67\textwidth]{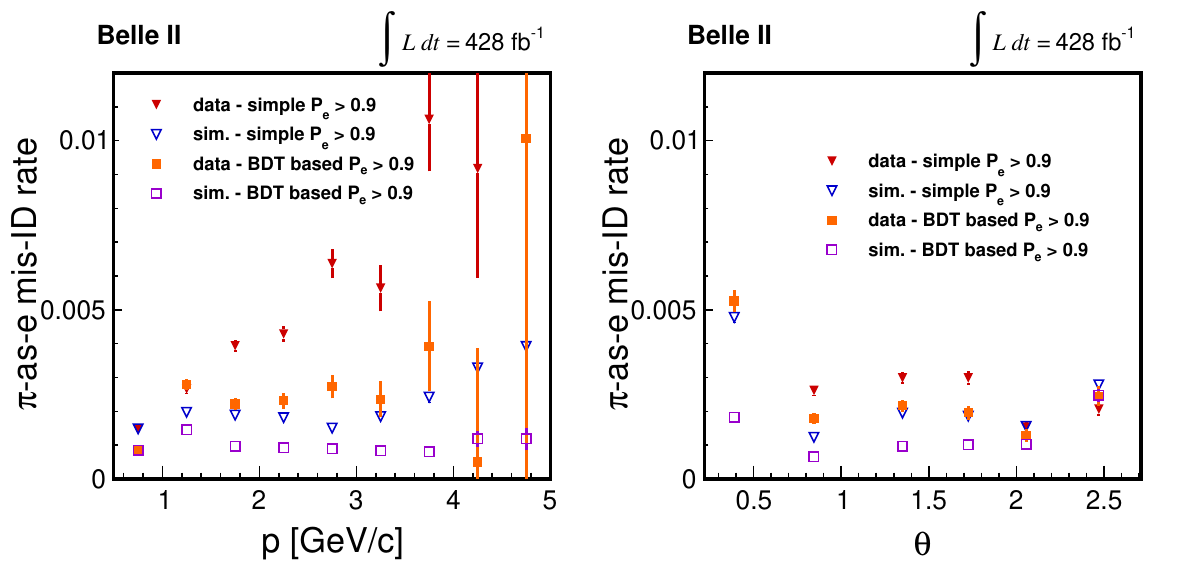} 
        \caption{$\pi$-as-$e$ mis-ID rate as a function of the momentum (left) and polar angle (right) as measured in the $D^*$ sample. Filled (empty) markers represent the data (simulation), with the downward pointing triangles showing the performance obtained with the simple $P_e > 0.9$ selection and the squares the BDT-based $P_e > 0.9$ selection.} 
        \label{fig:pi_to_e_fakes}
    \end{center}
\end{figure*}

Figure~\ref{fig:pi_to_e_fakes} shows the probability of mis-identifying a pion as an electron as a function of momentum and polar angle for both data and simulation, utilizing the simple probability and BDT-based probability selections. We see that in general the mis-ID rates measured in the data are substantially (a factor 2-3) higher than observed in the simulation, with the BDT-based selection giving better performance than the simple selection. The most critical region is the low-$\theta$ (forward) region, which is severely affected by the beam-related backgrounds.

\begin{figure*}[htbp]
    \begin{center}
        \includegraphics[width=0.67\textwidth]{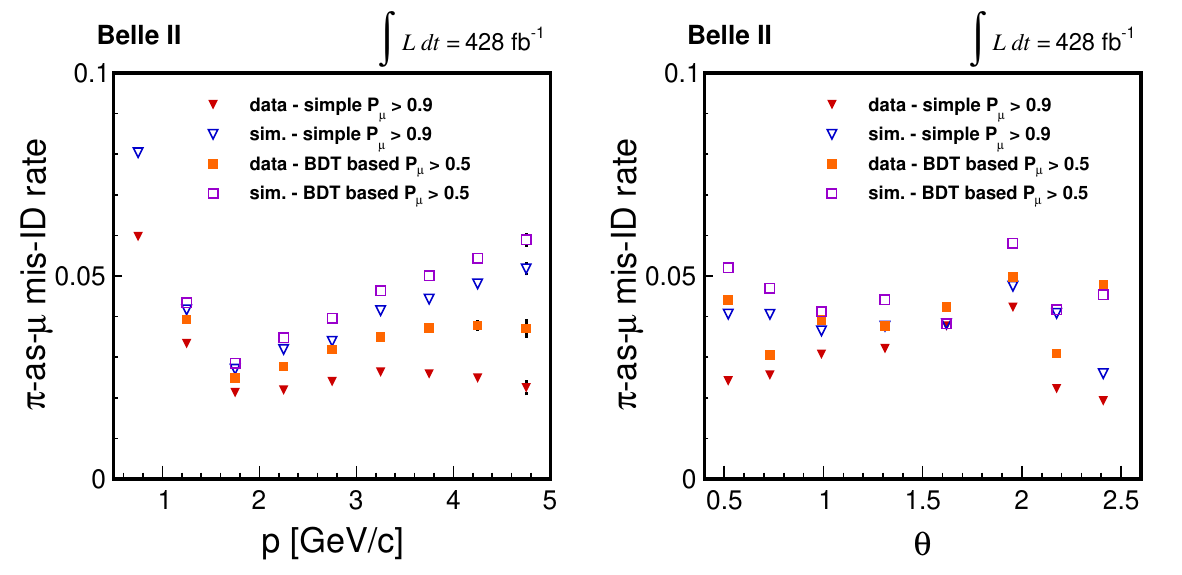} 
        \caption{$\pi$-as-$\mu$ mis-ID rate as a function of the momentum (left) and polar angle (right) as measured in the $D^*$ sample. Filled (empty) markers represent the data (simulation), with the downward pointing triangles showing the performance obtained with the simple $P_{\mu} > 0.9$ selection and the squares the BDT-based $P_{\mu} > 0.5$ selection.} 
        \label{fig:pi_to_mu_fakes}
    \end{center}
\end{figure*}

Figure~\ref{fig:pi_to_mu_fakes} shows that the probability for a pion to be mis-identified as a muon is actually lower in data than is simulation, particularly at higher momentum, with the simple probability providing better rejection compared to the BDT-based approach.  

\begin{figure*}[htbp]
    \begin{center}
        \includegraphics[width=0.67\textwidth]{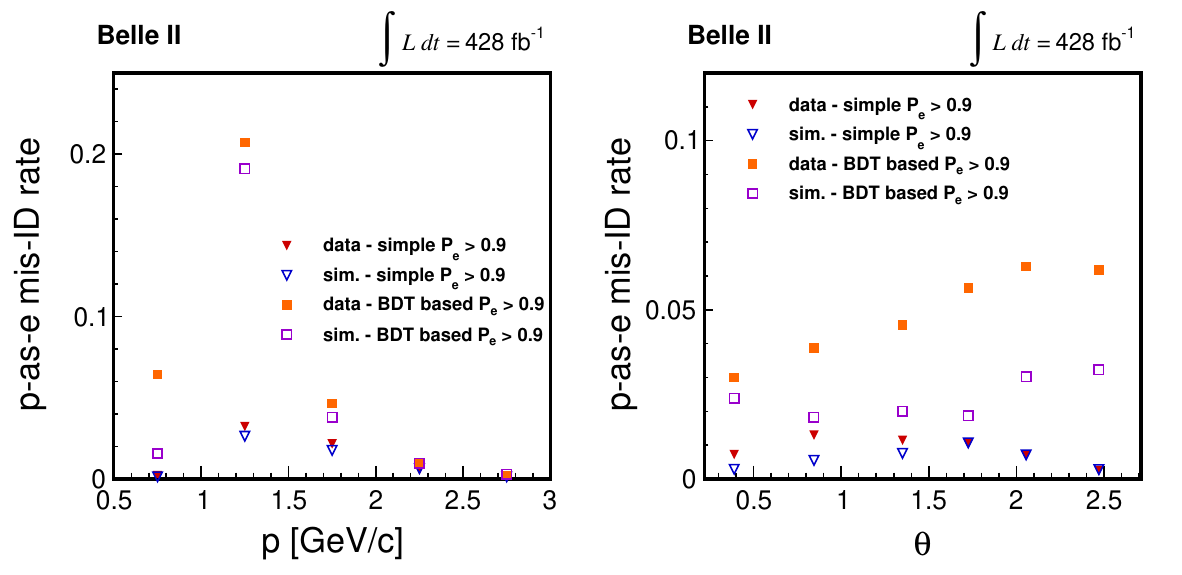} 
        \caption{$p$-as-$e$ mis-ID rate as a function of the momentum (left) and polar angle (right) as measured in the $\Lambda$ sample. Filled (empty) markers represent the data (simulation), with the downward pointing triangles showing the performance obtained with the simple $P_e > 0.9$ selection and the squares the BDT-based $P_e > 0.9$ selection.} 
        \label{fig:p_to_e_fakes}
    \end{center}
\end{figure*}

Figure~\ref{fig:p_to_e_fakes} shows the $p$-as-$e$ mis-ID rates; in general the simple probability performs significantly better than the BDT-based selection, especially for momenta between 1.0 and $1.5\,\gevc$, where the BDT-based $p$-as-$e$ mis-ID rate reaches 20\%. The poor performance of the BDT can be explained with the fact that this selection strategy has been optimized to discriminate leptons from pions, and this is particularly evident in the region $1.0 < p < 1.5\,\gevc$, where electrons and protons have similar \dEdx*. For this reason, the simple probability, which guarantees mis-ID rates below the few \% level, is preferred in analyses that require good (anti-)proton rejection. 

\subsection{Data/simulation correction tables}

To correct for data/simulation differences in physics analysis, we compile correction tables with standard binning in momentum and polar angle, separately for positive and negative charges. We consider four working points for the selection on the simple and BDT-based probabilities, requiring $P > 0.5, 0.9, 0.95$, and 0.99. In the subregions where the coverages of different control samples overlap, we take the weighted average of the results obtained in the individual control samples. The compatibility of the different results is good and we add the difference between the central values of the individual determinations as systematic uncertainty.
The statistical and systematic uncertainties of the correction tables are then propagated to the physics analyses.

In the following we show selected examples for $e^+$ and $e^-$, using the BDT-based probability, and for $\mu^+$ and $\mu^-$, using the simple probability. In all cases we show the results for the working point $P_{e/\mu} > 0.9$.

\begin{figure*}[htbp]
    \begin{center}
        \includegraphics[width=0.8\textwidth]{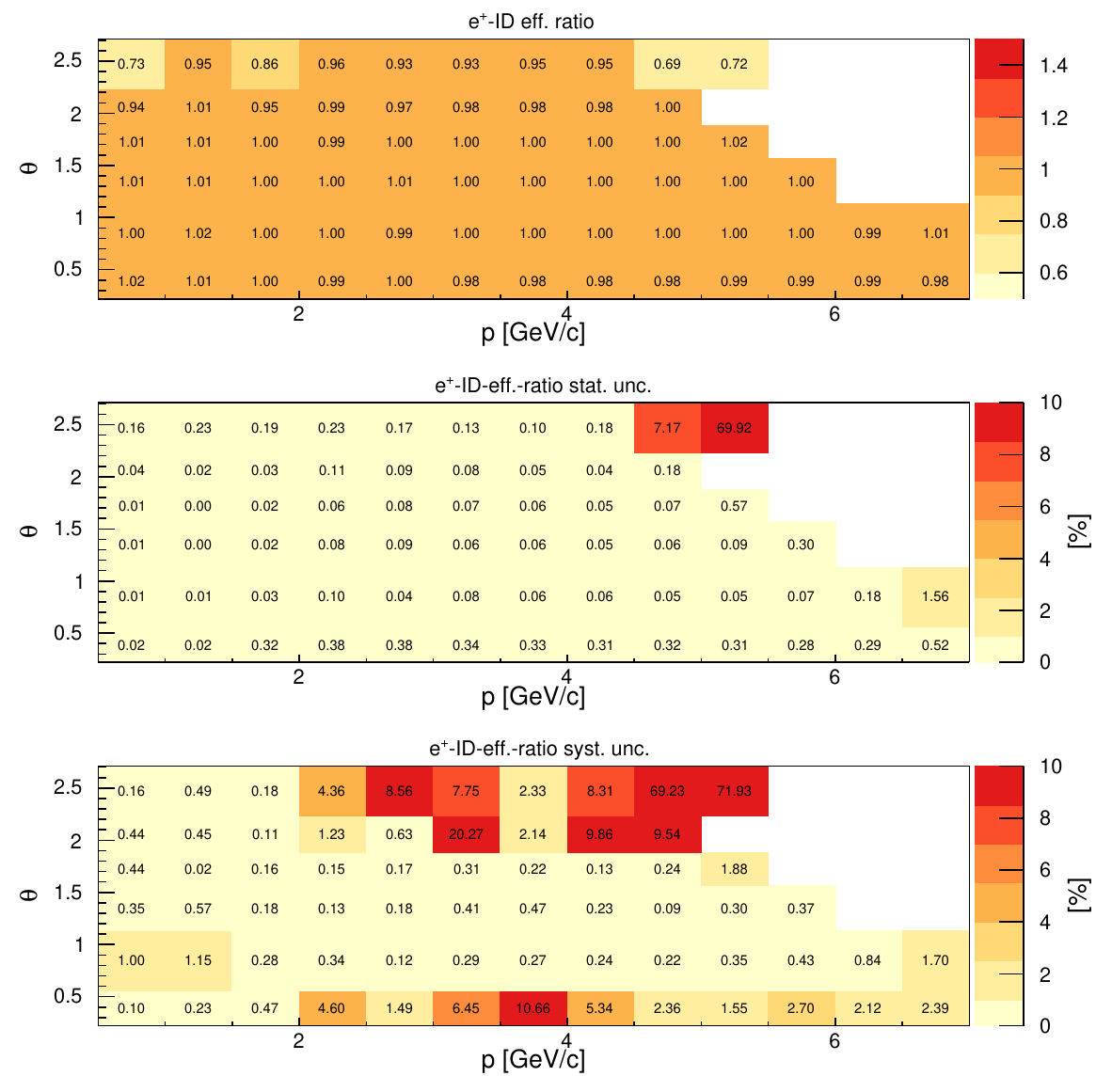} 
        \caption{Ratio of the $e^+$ ID efficiency in data to that in simulation (top) and its relative statistical (center) and systematic (bottom) uncertainties for the BDT-based $P_e > 0.9$ selection in subregions of momentum and polar angle.} 
        \label{fig:dataMC_e_plus}
    \end{center}
\end{figure*}

\begin{figure*}[htbp]
    \begin{center}
        \includegraphics[width=0.8\textwidth]{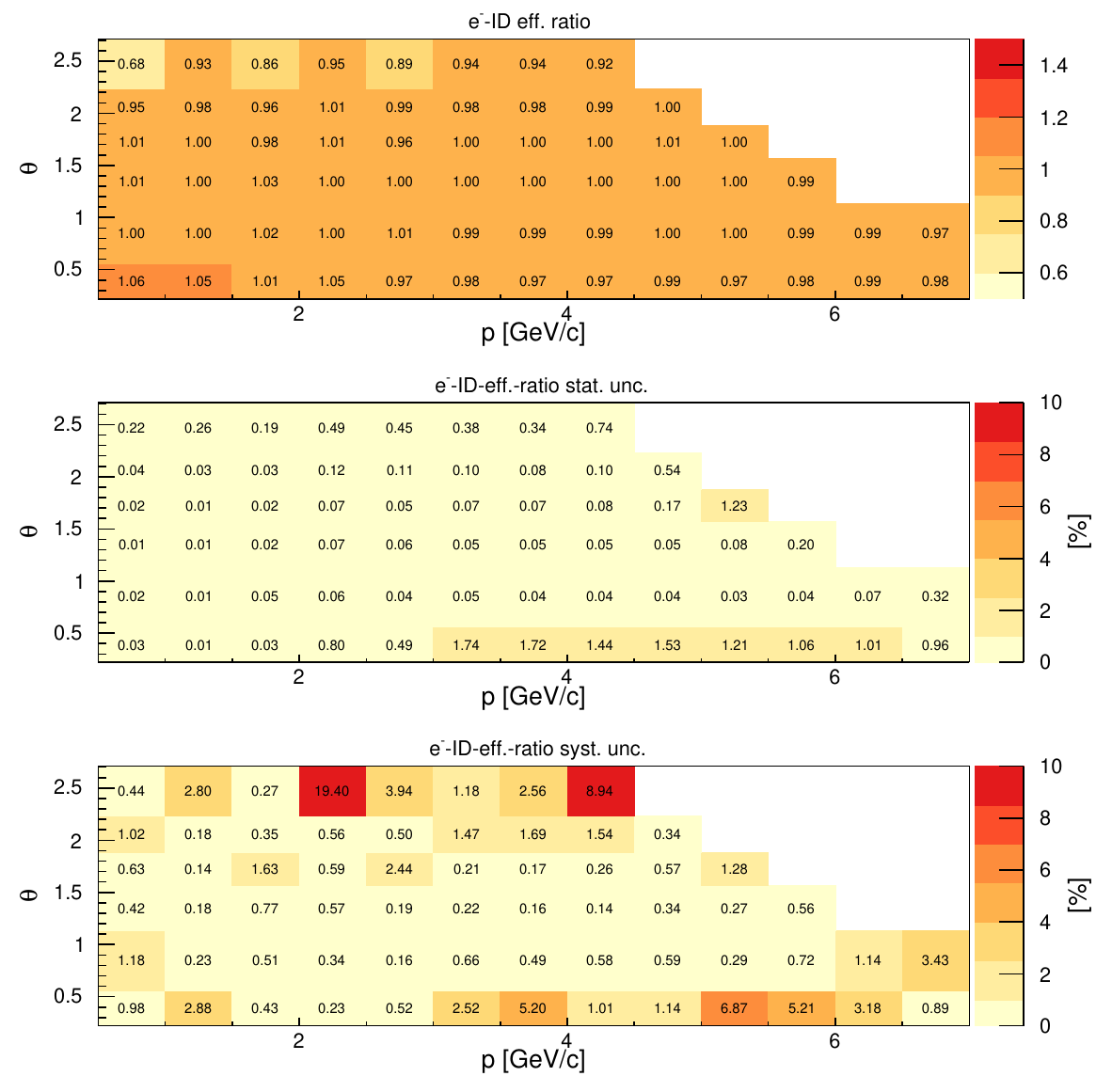} 
        \caption{Ratio of the $e^-$ ID efficiency in data to that in simulation (top) and its relative statistical (center) and systematic (bottom) uncertainties for the BDT-based $P_e > 0.9$ selection in subregions of momentum and polar angle.} 
        \label{fig:dataMC_e_minus}
    \end{center}
\end{figure*}

Figures~\ref{fig:dataMC_e_plus} and \ref{fig:dataMC_e_minus} show the ratios of electron ID efficiencies in data and simulation for $e^+$ and $e^-$, respectively. Overall the agreement between data and simulation is within 1\% in the barrel region of the detector, and within about 5\% in the forward and backward regions, except for individual cells where the deviation from unity can reach 30\%, however with a large uncertainty. Statistical uncertainties are below the 1 permille level in the bulk of the distribution, with the systematics being at the few permille level across most of the plane. Systematic uncertainties are large in some ($p$, $\theta$) subregions of the forward and backward endcaps, where backgrounds are poorly modeled and the agreement between different control samples is far from perfect.

\begin{figure*}[htbp]
    \begin{center}
        \includegraphics[width=0.8\textwidth]{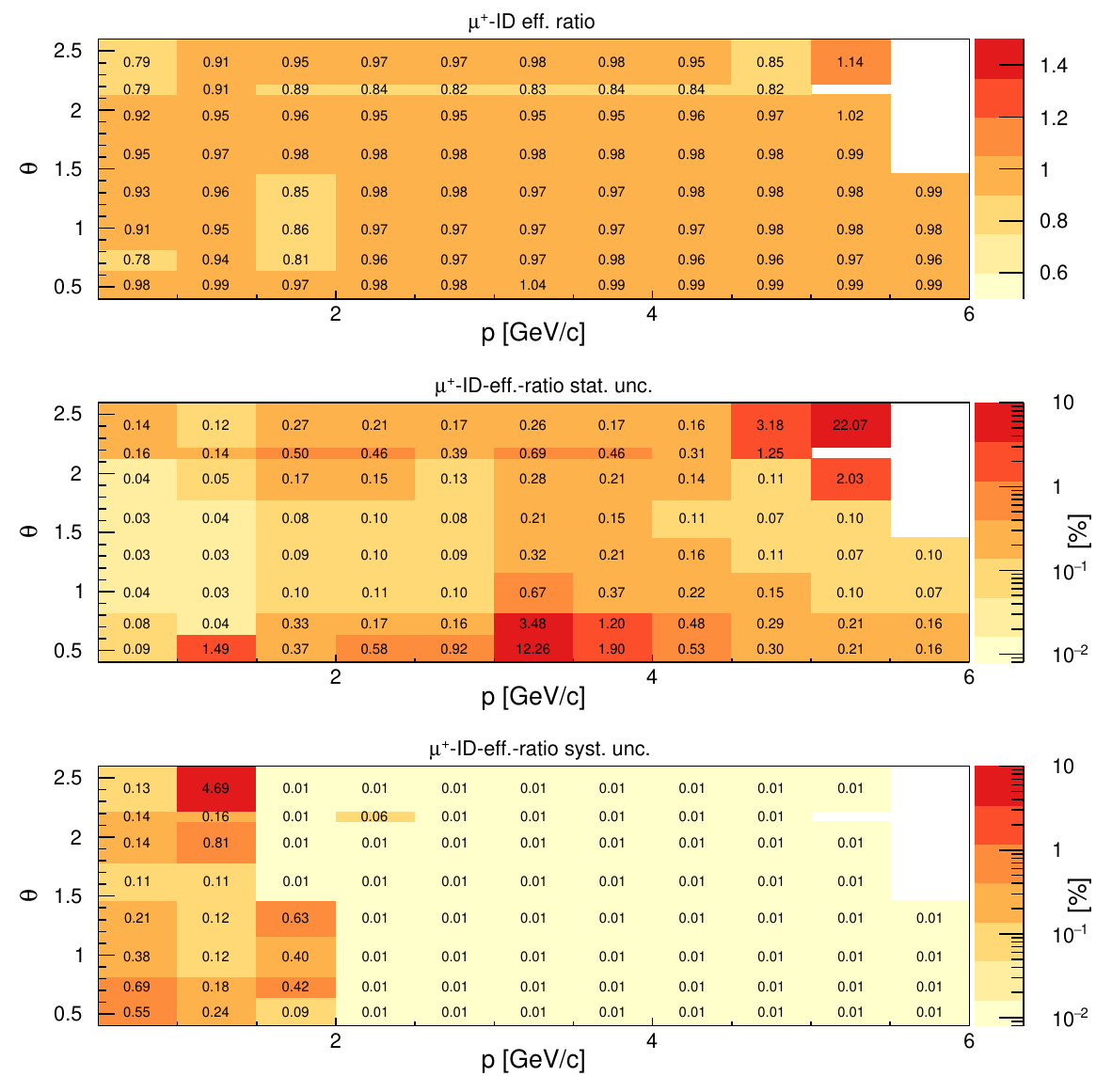} 
        \caption{Ratio of the $\mu^+$ ID efficiency in data to that in simulation (top) and its relative statistical (center) and systematic (bottom) uncertainties for the simple $P_{\mu} > 0.9$ selection in subregions of momentum and polar angle.
        We use a logarithmic color scale for the uncertainties.} 
        \label{fig:dataMC_mu_plus}
    \end{center}
\end{figure*}

\begin{figure*}[htbp]
    \begin{center}
        \includegraphics[width=0.8\textwidth]{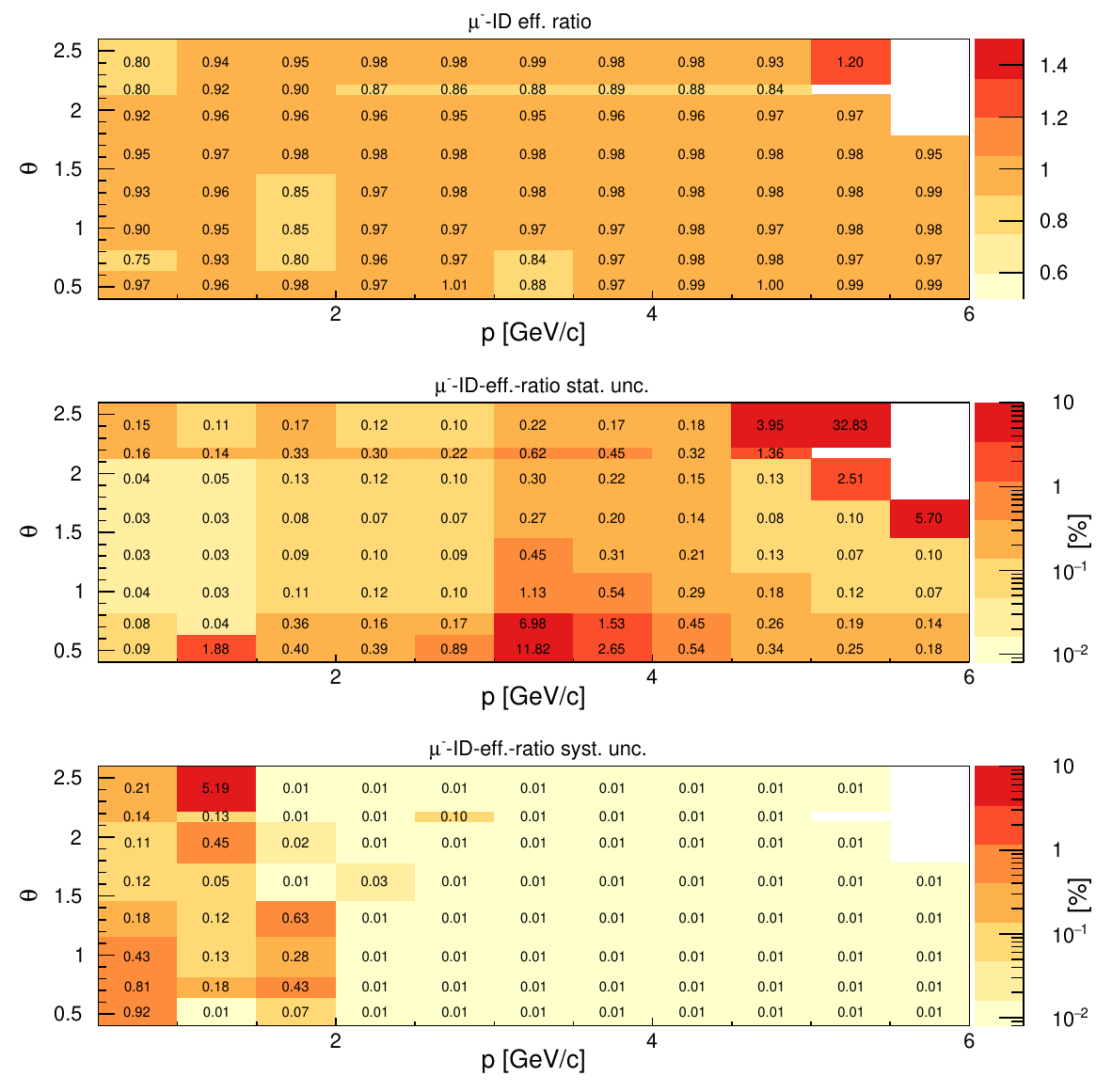} 
        \caption{Ratio of the $\mu^-$ ID efficiency in data to that in simulation (top) and its relative statistical (center) and systematic (bottom) uncertainties for the simple $P_{\mu} > 0.9$ selection in subregions of momentum and polar angle.
        We use a logarithmic color scale for the uncertainties.} 
        \label{fig:dataMC_mu_minus}
    \end{center}
\end{figure*}

Figures~\ref{fig:dataMC_mu_plus} and \ref{fig:dataMC_mu_minus} show the ratios of muon ID efficiencies in data and simulation for $\mu^+$ and $\mu^-$, respectively. On average, the simulation overestimates the performance seen in the data by a few $\%$. Large statistical uncertainties are observed in few ($p$, $\theta$) subregions where the coverage of our control samples is poor and the systematic uncertainties are typically much smaller than the statistical uncertainties.

\begin{figure*}[htbp]
    \begin{center}
        \includegraphics[width=0.8\textwidth]{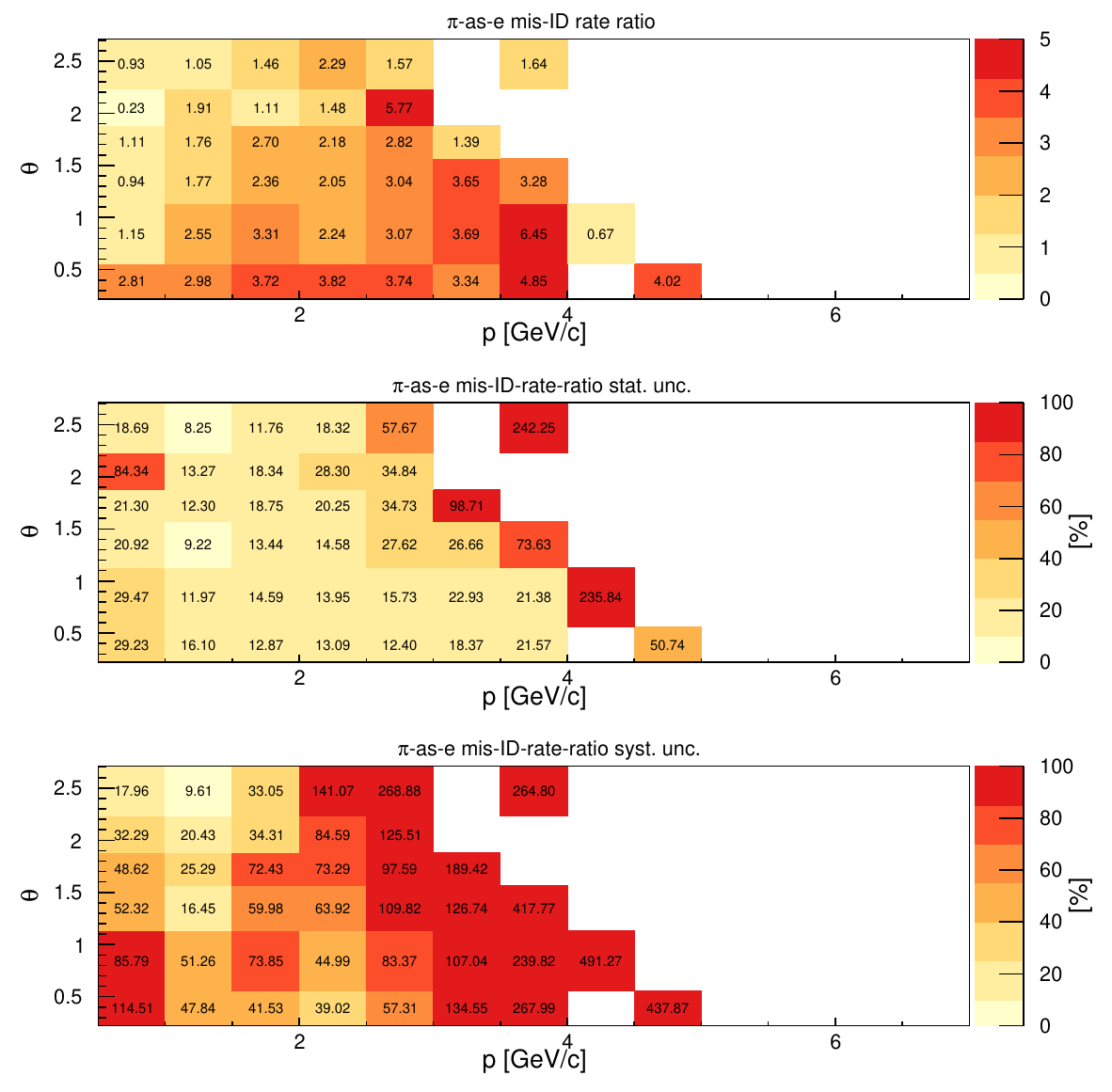}
        \caption{Ratio of the $\pi$-as-$e$ mis-ID rate in data to that in simulation (top) and its relative statistical (center) and systematic (bottom) uncertainties for the BDT-based $P_e > 0.9$ selection in subregions of momentum and polar angle.} 
        \label{fig:dataMC_pi_as_e}
    \end{center}
\end{figure*}

\begin{figure*}[htbp]
    \begin{center}
        \includegraphics[width=0.8\textwidth]{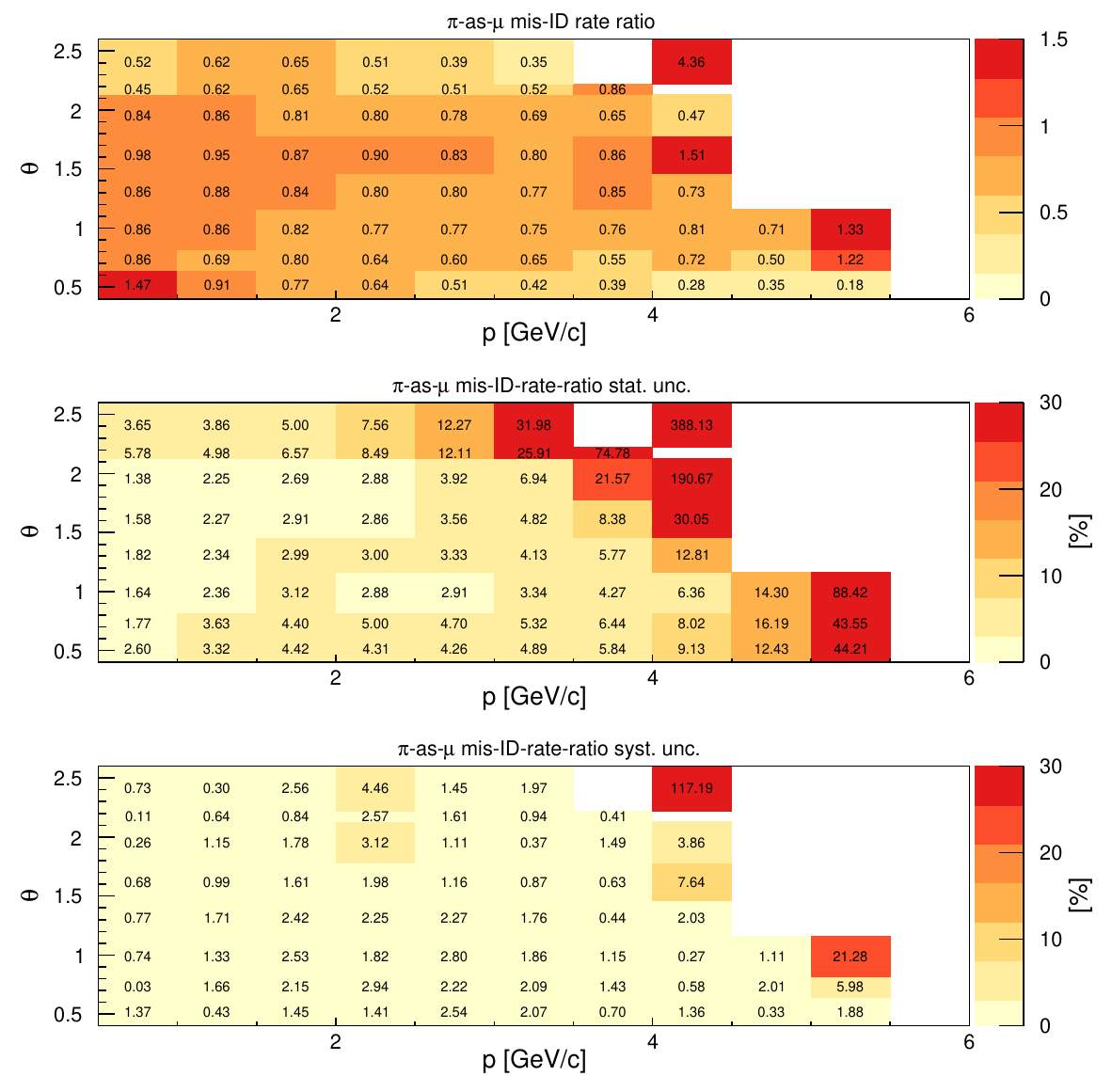}
        \caption{Ratio of the $\pi$-as-$\mu$ mis-ID rate in data to that in simulation (top) and its relative statistical (center) and systematic (bottom) uncertainties for the simple $P_{\mu} > 0.9$ selection in subregions of momentum and polar angle.} 
        \label{fig:dataMC_pi_as_mu}
    \end{center}
\end{figure*}

The pion-as-lepton mis-ID rates are studied using the pions originating from the $D^0$ decay in the $D^* \to D^0 \pi_{\rm soft}$ control sample. We use the same selection and binning as shown in the previous plots, but show the results combining positive and negative tracks. Figure~\ref{fig:dataMC_pi_as_e} (\ref{fig:dataMC_pi_as_mu}) shows the data/simulation ratios for the $\pi$-as-$e$ ($\pi$-as-$\mu$) mis-ID rate. The main limitation of this control sample is that the momentum coverage is limited to $4\,\gevc$. 

In the case of the electron identification, the simulation underestimates the mis-ID rate by a factor 2-3 across most of the plane. Both statistical and systematic uncertainties are typically between 30\% and 200\%: this is due to the fact that the mis-ID rate is very small, typically at the permille level. While the relative uncertainties are large, the overall impact on the physics results will be small, due to the small absolute values of the mis-ID rates.
For the muons, the simulation typically overestimates the mis-ID rate by 10--20\%. Relative statistical and systematic uncertainties are smaller than in the electron case, but in this case they refer to a quantity that is one order of magnitude larger.

\section{Conclusions and future prospects}
\label{sec:conclusions}

Lepton identification at \BelleII has proven to be effective in sharpening the physics reach of many analyses of the vast experimental program (see e.g. Refs.~\cite{Belle-II:2024vvr,Belle-II:2025yjp}).
A qualitative comparison with the performance achieved by \babar\ and Belle shows similar performance in electron ID, with efficiencies exceeding 95\% in most of the momentum range of interest and a $\pi$-as-$e$ mis-ID rate at the level of a few permille. The comparison of muon ID performance is more complicated, also because of its variations during the data-taking history of the first generation of $B$-Factory experiments. Nevertheless, all three experiments achieved a muon-ID efficiency above 90\%, with a $\pi$-as-$\mu$ mis-ID rate of about 5\% for momenta greater than $1\,\gevc$.

Despite the partial success of our first implementation of multivariate discrimination applied to muon ID, we believe that there is significant room for improvement, especially for muon identification for $p < 1\,\gevc$. Preliminary studies show that the extension of the neural network approach that was pioneered for $K$-vs-$\pi$ discrimination (see Ref.~\cite{Belle-II:2025tpe}) to leptons is advantageous with respect to the simple likelihood ratios. The unification of the strategies and software tools between lepton and hadron ID, which began as separate efforts, will simplify the maintenance of the code and ensure a uniform treatment of systematic uncertainties.

The loss of performance of the CDC \dEdx* due to the higher than anticipated injection backgrounds will be mitigated by a new calibration procedure that takes into account the time since the last bunch injection. Further improvements in PID might be possible through applications of machine learning to the reconstruction software of the TOP, ECL~\cite{Novosel:2023cki}, and KLM~\cite{Wang:2025jvi} subdetectors that are being attempted as we write.

%% file: acknowledgements-b2.tex
This work, based on data collected using the Belle II detector, which was built and commissioned prior to March 2019,
was supported by
Higher Education and Science Committee of the Republic of Armenia Grant No.~23LCG-1C011;
Australian Research Council and Research Grants
No.~DP200101792, 
No.~DP210101900, 
No.~DP210102831, 
No.~DE220100462, 
No.~LE210100098, 
and
No.~LE230100085; 
Austrian Federal Ministry of Education, Science and Research,
Austrian Science Fund (FWF) Grants
DOI:~10.55776/P34529,
DOI:~10.55776/J4731,
DOI:~10.55776/J4625,
DOI:~10.55776/M3153,
and
DOI:~10.55776/PAT1836324,
and
Horizon 2020 ERC Starting Grant No.~947006 ``InterLeptons'';
Natural Sciences and Engineering Research Council of Canada, Digital Research Alliance of Canada, and Canada Foundation for Innovation;
National Key R\&D Program of China under Contract No.~2024YFA1610503,
and
No.~2024YFA1610504
National Natural Science Foundation of China and Research Grants
No.~11575017,
No.~11761141009,
No.~11705209,
No.~11975076,
No.~12135005,
No.~12150004,
No.~12161141008,
No.~12405099,
No.~12475093,
and
No.~12175041,
and Shandong Provincial Natural Science Foundation Project~ZR2022JQ02;
the Czech Science Foundation Grant No. 22-18469S,  Regional funds of EU/MEYS: OPJAK
FORTE CZ.02.01.01/00/22\_008/0004632 
and
Charles University Grant Agency project No. 246122;
European Research Council, Seventh Framework PIEF-GA-2013-622527,
Horizon 2020 ERC-Advanced Grants No.~267104 and No.~884719,
Horizon 2020 ERC-Consolidator Grant No.~819127,
Horizon 2020 Marie Sklodowska-Curie Grant Agreement No.~700525 ``NIOBE''
and
No.~101026516,
and
Horizon Europe Marie Sklodowska-Curie Staff Exchange project JENNIFER3 Grant Agreement No.~101183137 (European grants);
L’Institut National de Physique Nucl\'eaire et de Physique des
Particules (IN2P3) du CNRS under Project Identification No.
CNRS-IN2P3-14-PP-033
and L’Agence Nationale de la Recherche (ANR) under Grant No. ANR-23-CE31-
0018 and ANR-25-CE31-1333 (France);
BMFTR, DFG, HGF, MPG, and AvH Foundation (Germany);
Department of Atomic Energy under Project Identification No.~RTI 4002,
Department of Science and Technology,
and
UPES SEED funding programs
No.~UPES/R\&D-SEED-INFRA/17052023/01 and
No.~UPES/R\&D-SOE/20062022/06 (India);
Israel Science Foundation Grant No.~2476/17,
U.S.-Israel Binational Science Foundation Grant No.~2016113, and
Israel Ministry of Science Grant No.~3-16543;
Istituto Nazionale di Fisica Nucleare and the Research Grants BELLE2,
and
the ICSC – Centro Nazionale di Ricerca in High Performance Computing, Big Data and Quantum Computing, funded by European Union – NextGenerationEU;
Japan Society for the Promotion of Science, Grant-in-Aid for Scientific Research Grants
No.~16H03993,
No.~16H06492,
No.~16K05323,
No.~17H01133,
No.~17H05405,
No.~18K03621,
No.~18H03710,
No.~18H05226,
No.~19H00682, 
No.~20H05850,
No.~20H05858,
No.~22H00144,
No.~22K14056,
No.~22K21347,
No.~23H05433,
No.~26220706,
No.~26400255,
and
No.~26H02056,
and
the Ministry of Education, Culture, Sports, Science, and Technology (MEXT) of Japan;  
National Research Foundation (NRF) of Korea Grants
No.~2021R1-F1A-1064008,
No.~2022R1-A2C-1003993,
No.~RS-2018-NR031074,
No.~RS-2021-NR060129,
No.~RS-2024-00354342,
No.~RS-2025-02219521,
No.~RS-2026-25471491,
No.~RS-2026-25480677,
and
No.~RS-2026-25486791,
Radiation Science Research Institute,
Foreign Large-Size Research Facility Application Supporting project,
the Global Science Experimental Data Hub Center, the Korea Institute of Science and
Technology Information (K26L1M2C3)
and
KREONET/GLORIAD;
Universiti Malaya RU grant, Akademi Sains Malaysia, and Ministry of Education Malaysia;
Frontiers of Science Program Contracts
No.~FOINS-296,
No.~CB-221329,
No.~CB-236394,
No.~CB-254409,
and
No.~CB-180023, and SEP-CINVESTAV Research Grant No.~237 (Mexico);
the Polish Ministry of Science and Higher Education and the National Science Center;
the Ministry of Science and Higher Education of the Russian Federation
and
the HSE University Basic Research Program, Moscow;
University of Tabuk Research Grants
No.~S-0256-1438 and No.~S-0280-1439 (Saudi Arabia);
Slovenian Research Agency and Research Grants
No.~J1-50010
and
No.~P1-0135;
Ikerbasque, Basque Foundation for Science,
State Agency for Research of the Spanish Ministry of Science and Innovation through Grant No. PID2022-136510NB-C33, Spain,
the Severo Ochoa project CEX2023-001292-S funded by MICIU/AEI, State Secretariat for
Telecommunications and Digital Infrastructure with reference
TSI-069100-2023-0012, State Agency for Research of the Spanish Ministry
of Science, Innovation and Universities through Grant No
PID2024-156645NB-C21;
The Knut and Alice Wallenberg Foundation (Sweden), Contracts No.~2021.0174, No.~2021.0299, and No.~2023.0315;
National Science and Technology Council,
and
Ministry of Education (Taiwan);
Thailand Center of Excellence in Physics;
TUBITAK ULAKBIM (Turkey);
National Research Foundation of Ukraine, Project No.~2020.02/0257,
and
Ministry of Education and Science of Ukraine;
the U.S. National Science Foundation and Research Grants
No.~PHY-1913789 
and
No.~PHY-2111604, 
and the U.S. Department of Energy and Research Awards
No.~DE-AC06-76RLO1830, 
No.~DE-SC0007983, 
No.~DE-SC0009824, 
No.~DE-SC0009973, 
No.~DE-SC0010007, 
No.~DE-SC0010073, 
No.~DE-SC0010118, 
No.~DE-SC0010504, 
No.~DE-SC0011784, 
No.~DE-SC0012704, 
No.~DE-SC0019230, 
No.~DE-SC0021616, 
No.~DE-SC0022350, 
No.~DE-SC0023470; 
and
the Vietnam Academy of Science and Technology (VAST) under Grant
No.~DL0000.05/26-27.

These acknowledgements are not to be interpreted as an endorsement of any statement made
by any of our institutes, funding agencies, governments, or their representatives.

We thank the SuperKEKB team for delivering high-luminosity collisions;
the KEK cryogenics group for the efficient operation of the detector solenoid magnet and IBBelle on site;
the KEK Computer Research Center for on-site computing support; the NII for SINET6 network support;
and the raw-data centers hosted by BNL, DESY, GridKa, IN2P3, INFN, 
and the University of Victoria.

%% file: belle2.bib
@article{Akai:2018mbz,
    author = "Akai, Kazunori and Furukawa, Kazuro and Koiso, Haruyo",
    collaboration = "SuperKEKB",
    title = "{SuperKEKB Collider}",
    eprint = "1809.01958",
    archivePrefix = "arXiv",
    primaryClass = "physics.acc-ph",
    doi = "10.1016/j.nima.2018.08.017",
    journal = "Nucl. Instrum. Meth. A",
    volume = "907",
    pages = "188",
    year = "2018"
}

@article{Belle-II:2018jsg,
    author = "Altmannshofer, W. and others",
    editor = "Kou, E. and Urquijo, P.",
    collaboration = "Belle II Collaboration",
    title = "{The Belle II Physics Book}",
    eprint = "1808.10567",
    archivePrefix = "arXiv",
    primaryClass = "hep-ex",
    reportNumber = "KEK Preprint 2018-27, BELLE2-PUB-PH-2018-001, FERMILAB-PUB-18-398-T, JLAB-THY-18-2780, INT-PUB-18-047, UWThPh 2018-26",
    doi = "10.1093/ptep/ptz106",
    journal = "PTEP",
    volume = "2019",
    number = "12",
    pages = "123C01",
    year = "2019",
    note = "[Erratum: PTEP \textbf{2020} (2020), 029201]"
}

@article{Bevan:2014iga,
      author         = "{Ed.~A.~J.~Bevan, B.~Golob, Th.~Mannel, S.~Prell, and B.~D.~Yabsley}",
      title          = "{The physics of the $B$ Factories}",
      journal        = "Eur. Phys. J.",
      volume         = "C74",
      year           = "2014",
      pages          = "3026",
      doi            = "10.1140/epjc/s10052-014-3026-9",
      eprint         = "1406.6311",
      archivePrefix  = "arXiv",
      primaryClass   = "hep-ex",
      reportNumber   = "SLAC-PUB-15968, KEK-PREPRINT-2014-3,
                        FERMILAB-PUB-14-262-T",
      SLACcitation   = "%%CITATION = ARXIV:1406.6311;%%"
}

@article{Abe:2010sj,
      author         = "Abe, T. and others",
      title          = "{Belle II Technical Design Report}",
      collaboration  = "Belle II Collaboration",
      year           = "2010",
      eprint         = "1011.0352",
      archivePrefix  = "arXiv",
      primaryClass   = "physics.ins-det",
      reportNumber   = "KEK-REPORT-2010-1",
      SLACcitation   = "%%CITATION = ARXIV:1011.0352;%%",
}

@article{Belle-II:2021zvj,
    author = "Abudin\'en, F. and others",
    collaboration = "Belle II Collaboration",
    title = "{B-flavor tagging at Belle II}",
    eprint = "2110.00790",
    archivePrefix = "arXiv",
    primaryClass = "hep-ex",
    reportNumber = "BELLE2-PUB-TE-2021-002",
    doi = "10.1140/epjc/s10052-022-10180-9",
    journal = "Eur. Phys. J. C",
    volume = "82",
    number = "4",
    pages = "283",
    year = "2022"
}

@article{Keck:2018lcd,
    author = "Keck, T. and others",
    title = "{The Full Event Interpretation}: {An Exclusive Tagging Algorithm for the Belle II Experiment}",
    eprint = "1807.08680",
    archivePrefix = "arXiv",
    primaryClass = "hep-ex",
    reportNumber = "DESY-19-048",
    doi = "10.1007/s41781-019-0021-8",
    journal = "Comput. Softw. Big Sci.",
    volume = "3",
    number = "1",
    pages = "6",
    year = "2019"
}

@article{Sjostrand:2014zea,
      author         = "Sjöstrand, Torbjörn and Ask, Stefan and Christiansen,
                        Jesper R. and Corke, Richard and Desai, Nishita and Ilten,
                        Philip and Mrenna, Stephen and Prestel, Stefan and
                        Rasmussen, Christine O. and Skands, Peter Z.",
      title          = "{An Introduction to PYTHIA 8.2}",
      journal        = "Comput. Phys. Commun.",
      volume         = "191",
      year           = "2015",
      pages          = "159",
      doi            = "10.1016/j.cpc.2015.01.024",
      eprint         = "1410.3012",
      archivePrefix  = "arXiv",
      primaryClass   = "hep-ph",
      reportNumber   = "LU-TP-14-36, MCNET-14-22, CERN-PH-TH-2014-190,
                        FERMILAB-PUB-14-316-CD, DESY-14-178, SLAC-PUB-16122",
      SLACcitation   = "%%CITATION = ARXIV:1410.3012;%%"
}

@article{Lange:2001uf,
      author         = "Lange, D. J.",
      title          = "{The EvtGen particle decay simulation package}",
      booktitle      = "{Proceedings, 7th International Conference on B physics
                        at hadron machines (BEAUTY 2000): Maagan, Israel,
                        September 13-18, 2000}",
      journal        = "Nucl. Instrum. Meth. A",
      volume         = "462",
      year           = "2001",
      pages          = "152",
      doi            = "10.1016/S0168-9002(01)00089-4",
      SLACcitation   = "%%CITATION = NUIMA,A462,152;%%"
}

@article{Pivk:2004ty,
    author = "Pivk, Muriel and Le Diberder, Francois R.",
    title = "{SPlot: A Statistical tool to unfold data distributions}",
    eprint = "physics/0402083",
    archivePrefix = "arXiv",
    reportNumber = "LAL-04-07",
    doi = "10.1016/j.nima.2005.08.106",
    journal = "Nucl. Instrum. Meth. A",
    volume = "555",
    pages = "356--369",
    year = "2005"
}

@article{Staric:2008,
    author = "Stari{\v c}, M. and others",
    title = "{Likelihood analysis of patterns in a time-of-propagation (TOP) counter}",
    doi = "10.1016/j.nima.2008.07.018",
    journal = "Nucl. Instrum. Meth. A",
    volume = "595",
    year = "2008",
    pages = "252-255"
}

@article{Kuhr:2018lps,
    author = "Kuhr, T. and Pulvermacher, C. and Ritter, M. and Hauth, T. and Braun, N.",
    collaboration = "Belle II Framework Software Group",
    title = "{The Belle II Core Software}",
    eprint = "1809.04299",
    archivePrefix = "arXiv",
    primaryClass = "physics.comp-ph",
    doi = "10.1007/s41781-018-0017-9",
    journal = "Comput. Softw. Big Sci.",
    volume = "3",
    number = "1",
    pages = "1",
    year = "2019"
}

@misc{basf2-zenodo,
    author = "{Belle II Collaboration}",
    title = "{Belle II Analysis Software Framework (basf2)}",
    howpublished = "\url{https://doi.org/10.5281/zenodo.5574115}"
}

@article{Belle-II:2024lwr,
    author = "Adachi, I. and others",
    collaboration = "Belle II Collaboration",
    title = "{New graph-neural-network flavor tagger for Belle II and measurement of {\ensuremath{\sin 2\phi_1}} in {\ensuremath{B^0}}{\textrightarrow}J/{\ensuremath{\psi}}{\ensuremath{K_S^0}} decays}",
    eprint = "2402.17260",
    archivePrefix = "arXiv",
    primaryClass = "hep-ex",
    reportNumber = "Belle II Preprint 2024-006, KEK Preprint 2023-53",
    doi = "10.1103/PhysRevD.110.012001",
    journal = "Phys. Rev. D",
    volume = "110",
    number = "1",
    pages = "012001",
    year = "2024"
}

@article{Cranmer:2000du,
    author = "Cranmer, Kyle S.",
    title = "{Kernel estimation in high-energy physics}",
    eprint = "hep-ex/0011057",
    archivePrefix = "arXiv",
    doi = "10.1016/S0010-4655(00)00243-5",
    journal = "Comput. Phys. Commun.",
    volume = "136",
    pages = "198",
    year = "2001"
}

@article{Belle-II:2023vra,
    author = "Adachi, I. and others",
    collaboration = "Belle II Collaboration",
    title = "{Novel method for the identification of the production flavor of neutral charmed mesons}",
    eprint = "2304.02042",
    archivePrefix = "arXiv",
    primaryClass = "hep-ex",
    reportNumber = "Belle II Preprint 2023-005, KEK Preprint 2023-1",
    doi = "10.1103/PhysRevD.107.112010",
    journal = "Phys. Rev. D",
    volume = "107",
    number = "11",
    pages = "112010",
    year = "2023"
}

@article{GEANT4:2002zbu,
    author = "Agostinelli, S. and others",
    collaboration = "GEANT4 Collaboration",
    title = "{GEANT4--a simulation toolkit}",
    reportNumber = "SLAC-PUB-9350, FERMILAB-PUB-03-339, CERN-IT-2002-003",
    doi = "10.1016/S0168-9002(03)01368-8",
    journal = "Nucl. Instrum. Meth. A",
    volume = "506",
    pages = "250--303",
    year = "2003"
}

@article{Jadach:1999vf,
    author = "Jadach, S. and Ward, B. F. L. and Was, Z.",
    title = "{The Precision Monte Carlo event generator K K for two fermion final states in $e^+ e^-$ collisions}",
    eprint = "hep-ph/9912214",
    archivePrefix = "arXiv",
    reportNumber = "DESY-99-106, CERN-TH-99-235, UTHEP-99-08-01",
    doi = "10.1016/S0010-4655(00)00048-5",
    journal = "Comput. Phys. Commun.",
    volume = "130",
    pages = "260",
    year = "2000"
}

@article{Davidson:2010rw,
    author = "Davidson, N. and Nanava, G. and Przedzinski, T. and Richter-Was, E. and Was, Z.",
    title = "{Universal Interface of TAUOLA Technical and Physics Documentation}",
    eprint = "1002.0543",
    archivePrefix = "arXiv",
    primaryClass = "hep-ph",
    reportNumber = "IFJPAN-IV-2009-10",
    doi = "10.1016/j.cpc.2011.12.009",
    journal = "Comput. Phys. Commun.",
    volume = "183",
    pages = "821",
    year = "2012"
}

@article{Balossini:2008xr,
    author = "Balossini, G. and Bignamini, C. and Calame, C. M. Carloni and Montagna, G. and Nicrosini, O. and Piccinini, F.",
    title = "{Photon pair production at flavour factories with per mille accuracy}",
    eprint = "0801.3360",
    archivePrefix = "arXiv",
    primaryClass = "hep-ph",
    reportNumber = "FNT-T-2008-01, SHEP-08-05",
    doi = "10.1016/j.physletb.2008.04.007",
    journal = "Phys. Lett. B",
    volume = "663",
    pages = "209--213",
    year = "2008"
}

@article{Berends:1984gf,
    author = "Berends, Frits A. and Daverveldt, P. H. and Kleiss, R.",
    title = "{Complete Lowest Order Calculations for Four Lepton Final States in electron-Positron Collisions}",
    reportNumber = "Print-84-1007 (LEIDEN)",
    doi = "10.1016/0550-3213(85)90541-3",
    journal = "Nucl. Phys. B",
    volume = "253",
    pages = "441--463",
    year = "1985"
}

@article{Chapter34_pdg,
    author = "Groom, D.~E. and Klein, S.~R.",
    title = "{Passage of Particles Through Matter, in Particle Data Group, Review of particle physics}",
    doi = "10.1103/PhysRevD.110.030001",
    journal = "Phys. Rev. D",
    volume = "110",
    number = "3",
    pages = "Chapter 34",
    year = "2024"
}

@article{MOSER201685,
    title = {The Belle II DEPFET pixel detector},
    journal = {Nucl. Instrum. Meth. A},     
    volume = {831},
    pages = {85-87},
    year = {2016},
    issn = {0168-9002},
    doi = {https://doi.org/10.1016/j.nima.2016.02.078},
    url = {https://www.sciencedirect.com/science/article/pii/S0168900216002540},
    author = {Hans-Günther Moser}
}

@article{Adamczyk_2022,
    author = "Adamczyk, K. and others",
    collaboration = "Belle II SVD Group",
    title = "{The design, construction, operation and performance of the Belle~II silicon vertex detector}",
    eprint = "2201.09824",
    archivePrefix = "arXiv",
    primaryClass = "physics.ins-det",
    doi = "10.1088/1748-0221/17/11/P11042",
    journal = "JINST",
    volume = "17",
    number = "11",
    pages = "P11042",
    year = "2022"
}

@article{Yonenaga:2020eby,
    author = "Yonenaga, M. and others",
    title = "{Performance evaluation of the aerogel RICH counter for the Belle II spectrometer using early beam collision data}",
    eprint = "2008.06251",
    archivePrefix = "arXiv",
    primaryClass = "physics.ins-det",
    doi = "10.1093/ptep/ptaa090",
    journal = "PTEP",
    volume = "2020",
    number = "9",
    pages = "093H01",
    year = "2020"
}

@article{Atmacan:2025jmh,
    author = "Atmacan, H. and others",
    title = "{The imaging Time-of-Propagation detector at Belle II}",
    eprint = "2504.19090",
    archivePrefix = "arXiv",
    primaryClass = "hep-ex",
    reportNumber = "UCHEP-25-01, University of Cincinnati preprint UCHEP-25-01",
    doi = "10.1016/j.nima.2025.170627",
    journal = "Nucl. Instrum. Meth. A",
    volume = "1080",
    pages = "170627",
    year = "2025"
}

@article{Belle-II:2025tpe,
    author = "Adachi, I. and others",
    collaboration = "Belle II Collaboration",
    title = "{Charged-hadron identification at Belle II}",
    eprint = "2506.04355",
    archivePrefix = "arXiv",
    primaryClass = "hep-ex",
    reportNumber = "Belle II Preprint 2025-016, KEK Preprint 2025-15",
    doi = "10.1140/epjc/s10052-025-14627-7",
    journal = "Eur. Phys. J. C",
    volume = "85",
    number = "11",
    pages = "1237",
    year = "2025"
}

@article{Milesi:2020esq,
    author = "Milesi, Marco and Tan, Justin and Urquijo, Phillip",
    editor = "Doglioni, C. and Kim, D. and Stewart, G. A. and Silvestris, L. and Jackson, P. and Kamleh, W.",
    title = "{Lepton identification in Belle II using observables from the electromagnetic calorimeter and precision trackers}",
    doi = "10.1051/epjconf/202024506023",
    journal = "EPJ Web Conf.",
    volume = "245",
    pages = "06023",
    year = "2020"
}

@article{Longo:2020zqt,
    author = "Longo, S. and others",
    title = "{CsI(Tl) pulse shape discrimination with the Belle II electromagnetic calorimeter as a novel method to improve particle identification at electron{\textendash}positron colliders}",
    eprint = "2007.09642",
    archivePrefix = "arXiv",
    primaryClass = "physics.ins-det",
    doi = "10.1016/j.nima.2020.164562",
    journal = "Nucl. Instrum. Meth. A",
    volume = "982",
    pages = "164562",
    year = "2020"
}

@phdthesis{Gaiser:Phd,
    author = "Gaiser, John",
    title = "{Charmonium spectroscopy from radiative decays of the $J/\psi$ and $\psi^\prime$}",
    reportNumber = "SLAC-R-255",
    school = "Stanford University",
    year = "1982"
}

@phdthesis{Skwarnicki:1986xj,
    author = "Skwarnicki, Tomasz",
    title = "{A study of the radiative CASCADE transitions between the Upsilon-Prime and Upsilon resonances}",
    reportNumber = "DESY-F31-86-02, DESY-F-31-86-02",
    school = "Cracow, INP",
    year = "1986"
}

@article{Fox:1978vu,
    author = "Fox, Geoffrey C. and Wolfram, Stephen",
    title = "{Observables for the Analysis of Event Shapes in {\ensuremath{e^+ e^-}} Annihilation and Other Processes}",
    reportNumber = "CALT-68-680",
    doi = "10.1103/PhysRevLett.41.1581",
    journal = "Phys. Rev. Lett.",
    volume = "41",
    pages = "1581",
    year = "1978"
}

@article{Kuzmin:2020new,
    author = "Kuzmin, A.",
    editor = "Krammer, Manfred and Bergauer, Thomas and Dragicevic, Marko and Friedl, Markus and Jeitler, Manfred and Schieck, Jochen and Schwanda, Christoph",
    collaboration = "Belle II Calorimeter Team",
    title = "{Electromagnetic calorimeter of Belle II}",
    doi = "10.1016/j.nima.2019.05.076",
    journal = "Nucl. Instrum. Meth. A",
    volume = "958",
    pages = "162235",
    year = "2020"
}

@article{Ketter:2025yqm,
    author = "Ketter, C. and others",
    title = "{Design and commissioning of readout electronics for a {\ensuremath{K_L^0}} and {\ensuremath{\mu}} detector at the Belle II experiment}",
    doi = "10.1016/j.nima.2025.170893",
    journal = "Nucl. Instrum. Meth. A",
    volume = "1082",
    pages = "170893",
    year = "2026"
}

@article{Belle-II:2024vvr,
    author = "Adachi, Ichiro and others",
    collaboration = "Belle II Collaboration",
    title = "{Test of light-lepton universality in {\ensuremath{\tau}} decays with the Belle II experiment}",
    doi = "10.1007/JHEP08(2024)205",
    journal = "JHEP",
    volume = "08",
    pages = "205",
    year = "2024"
}

@article{Belle-II:2025yjp,
    author = "Adachi, I. and others",
    collaboration = "Belle II Collaboration",
    title = "{Test of lepton flavor universality with measurements of $R(D^+)$ and $R(D^{*+})$ using semileptonic B tagging at the Belle II experiment}",
    doi = "10.1103/fmn3-h8fy",
    journal = "Phys. Rev. D",
    volume = "112",
    pages = "032010",
    year = "2025"
}

@article{Wang:2025jvi,
    author = "Wang, Zihan and others",
    title = "{Muon identification with Deep Neural Network in the Belle II K-Long and Muon detector}",
    eprint = "2503.11351",
    archivePrefix = "arXiv",
    primaryClass = "hep-ex",
    doi = "10.1016/j.nima.2025.170814",
    journal = "Nucl. Instrum. Meth. A",
    volume = "1081",
    pages = "170814",
    year = "2026"
}

@article{Novosel:2023cki,
    author = "Novosel, Anja and Narimani Charan, Abtin and {\v{S}}antelj, Luka and Ferber, Torben and Kri{\v{z}}an, Peter and Golob, Bo{\v{s}}tjan",
    title = "{Identification of light leptons and pions in the electromagnetic calorimeter of Belle II}",
    eprint = "2301.05074",
    archivePrefix = "arXiv",
    primaryClass = "hep-ex",
    doi = "10.1016/j.nima.2023.168630",
    journal = "Nucl. Instrum. Meth. A",
    volume = "1056",
    pages = "168630",
    year = "2023"
}
